\documentclass[3p,times]{elsarticle}
\usepackage{graphicx} % Required for inserting images
\usepackage{siunitx}
\usepackage{xurl}
\usepackage{orcidlink}
\usepackage{amsmath}
\usepackage{bm}
\usepackage{comment}
\usepackage[version=4]{mhchem}
\newcommand{\bfL}{\textbf{L}}
\newcommand{\bfy}{\textbf{y}}
\newcommand{\bff}{\textbf{f}}

\newcommand{\tgz}{\hat{z}}
\newcommand{\an}{\alpha_{\min}}
\newcommand{\ax}{\alpha_{\max}}

\begin{document}
\begin{frontmatter}
\title{Tangent point position and atmosphere composition in limb scanning instruments}

\author[1]{Francesco Pio De Cosmo\orcidlink{0009-0008-2780-7572}}
\ead{francescopiodecosmo@cnr.it}
\author[1]{Luca Sgheri \orcidlink{0000-0001-5734-0856}\thanks{Corresponding author}}
\ead{luca.sgheri@cnr.it}

\affiliation[1] {organization={CNR-IAC},
                 addressline={Via Madonna del Piano, 10},
                 city={Sesto Fiorentino},
                 postcode={I-50019},
                 state={FI},
                 country={Italy}
}

\begin{abstract}
In the forward model for limb-scanning instruments, ray tracing must be accounted for because variations in air refractivity cause the lines of sight to bend from straight paths into curves. The tangent point of a line of sight is defined as the minimum height point. The lines of sight can be adjusted by varying the nadir angles of the instrument, which must be calibrated to account for Earth's ellipticity and the actual atmospheric conditions sampled by these paths. In this study, we investigate the relationship between the nadir angles, atmospheric properties, and tangent point positions, using a configuration relevant to the CAIRT instrument. Atmospheric data from reanalysis databases are utilized, and the lines of sight are determined by numerically solving the Eikonal equation. Our findings are compared to those of a previous study conducted for the MIPAS instrument, highlighting key differences.
\end{abstract}

\end{frontmatter}

\section{Introduction}
Limb scanning is an observation geometry in which the instrument's lines of sight penetrate the atmosphere, but do not reach Earth surface. A limb scanning instrument has in principle a much better resolving power with respect to a nadir looking instrument. For instance the spectra obtained from the rays sampling the upper part of the atmosphere only depend on the low density atmospheric layers. These spectra are used to resolve the upper atmosphere composition. In the nadir looking instrument, the sensitivity to changes in the upper layers composition is very low, because the information is masked by the much stronger signals coming from the lower layers.

However, limb instruments faces two problems. First, the mechanics of the instrument is normally more complicated with respect to a nadir looking instrument. Second, also from the point of view of the retrieval algorithm, there is the additional difficulty of predicting the shape of the lines of sight, since the changes in the atmosphere refractive index bend the rays from straight lines into curves.

Thus, not many limb scanning instruments operated in the past. The payload of the ESA ENVISAT satellite, that was active from 2002 to 2012, contained two limb-scanning instruments, SCIAMACHY (Scanning Imaging Absorption Spectrometer for Atmospheric Cartography) \cite{burrows1995} and MIPAS (Michelson Interferometer for Passive Atmospheric Sounding) \cite{fischer2008}. In particular, in the last operative reprocessing \cite{dinelli2021,raspollini2022} MIPAS obtained geolocated profiles of tangent pressures, temperature and $20$ atmospheric gases on a vertical grid ranging from $6$ to $60$ kilometers.

The ACE-FTS (Atmospheric Chemistry Experiment – Fourier Transform Spectrometer) is a limb-scanning satellite instrument aboard the Canadian SCISAT mission that measures atmospheric composition through solar occultation. It provides high-resolution infrared spectra used to retrieve vertical profiles of trace gases, temperature, and pressure in the stratosphere and upper troposphere \cite{boone2020}.

The MLS (Microwave Limb Sounder) on NASA's Aura satellite measures thermal emission from the limb of Earth's atmosphere in the microwave spectrum, enabling the retrieval of vertical profiles of atmospheric gases, temperature, and cloud ice. MLS data are essential for studying stratospheric chemistry and dynamics \cite{waters2022}.

ACE-FTS and MLS are instrument launched respectively in 2003 and 2004. They successfully measured the atmosphere composition for more than twenty years, but unfortunately they are close to the end of their operative life.

The CAIRT (Changing-Atmosphere Infra-Red Tomography Explorer) project \cite{riese2023,fischer2024,rhode2024} is a new limb scanning instrument project that may be seen as a MIPAS successor. At the moment (05/2025) the project has the status of Earth Explorer 11 candidate. The final selection will take place in Summer 2025. The CAIRT instrument has outstanding 3D retrieval capabilities, which may substantially contribute to the monitoring of the atmosphere, including the mesosphere. There are however some challenges, typical of a limb scanning observational geometry, that need to be tackled.

In the forward model for limb-scanning instruments, ray-tracing must be taken into account, as variations in atmospheric refractivity cause the lines of sight to bend from straight paths into curves. The shape of these curved paths depends on several factors, including the refractive index model, the assumed atmospheric composition and of course the ray-tracing algorithm. A detailed analysis of these dependencies has already been conducted for the MIPAS instrument \cite{ridolfi2014}.
%One of the most important issue is ray-tracing, that is essential to locate the spectra acquired by the instrument.
In this paper, we will revisit this study and highlight the differences by providing a more accurate analysis of the tangent point location as a function of the atmosphere. The study is conducted within the framework of the CAIRT Analysis and Synergy with IASI-NG (CASIA) project of the Italian Space Agency (ASI).

In Section ~\ref{se2:background}, we provide an overview of the key components involved in the ray-tracing procedure. This includes a description of the satellite orbit and atmospheric grid, the geometry of the lines of sight, the role of refractivity, the atmospheric models adopted in the simulations, and the instrumental settings for both MIPAS and CAIRT. Section ~\ref{se3:Results} presents the main results of our analysis, focusing on the determination of the nadir angles and the accuracy of the prediction of the tangent point. In Section~\ref{se4:MIPAS-CAIRT}, we compare the ray-tracing outputs obtained using MIPAS and CAIRT settings, highlighting the differences arising from their distinct horizontal resolutions. In the Appendix, we report the main issues in the numerical solution of the Eikonal equation.

\section{Background and instrumental settings}\label{se2:background}
\subsection{Satellite orbit and atmospheric grid}
We start by briefly reviewing the fundamental concepts of satellite observation. General references on this topic include \cite{campbell2011,vallado2013}. A common and reasonable assumption is that the satellite's orbit around Earth can be approximated as circular. This approximation is justified by the requirement for the satellite to maintain an almost constant altitude above the Earth's surface \cite{esa_polar_sun_sync}, so the orbit eccentricity must be low. 

Most remote sensing instruments, including CAIRT, are carried on satellites in sun-synchronous orbits. A sun-synchronous orbit ensures that the satellite passes over a given point on Earth's surface at nearly the same local solar time each day. In such orbits, the inclination $\theta$---defined as the angle between the equatorial plane and the orbital plane---is retrograde, meaning $\theta > 90^\circ$. Moreover, the orbital radius $r$ and the inclination $\theta$ are interdependent: the precession of the orbital plane caused by Earth's oblateness must precisely offset the precession resulting from Earth's annual revolution around the Sun.

For the Earth shape, we assume the WGS84 geoid model \cite{wgs84}. From the mathematical point of view the WGS84 model assumes that Earth is a rotational ellipsoid with an equatorial radius (semi major axis) $R_e=6378.137$ \si{km} and a polar radius (semi minor axis) $R_p=6356.752$ \si{km}. 
In this geometry, the section of Earth on the orbit plane is an ellipse with semi major axis $R_e$ and semi minor axis defined  by the following formula
\begin{equation}\label{eq:rtheta}
R_\theta=\sqrt{\frac{R_p^2R_e^2(1+\tan^2(\theta))}{R_p^2+R_e^2\tan^2(\theta)}}.
\end{equation}
which depends on the inclination angle $\theta$. 

In the orbital plane, positions on the Earth's surface can be represented using Cartesian coordinates with the origin at the Earth's center. In this coordinate system, the $x$-axis lies along the intersection of the orbital plane with the equatorial plane, while the $y$-axis is orthogonal to the $x$-axis and points northward within the orbital plane.
%In the orbit plane, the equation of Earth's surface is the ellipse given by the following parametric equations:
The Earth's surface, projected into this plane, can be approximated by a parametric ellipse described by the following equations:
\begin{equation}\label{eq:ellipse}
 \left\{\begin{aligned}
x(t)&=R_e\cos(t)\\
y(t)&=R_\theta \sin(t)
\end{aligned}\right.
\end{equation}
where the parameter $t \in [0,2 \pi]$ indicates the parametric angle along the ellipse in the orbital plane. The latitude and longitude of Earth's surface points can be determined from this angle. 
\begin{figure}[h!] 
    \centering
    \includegraphics[width=0.4\linewidth]{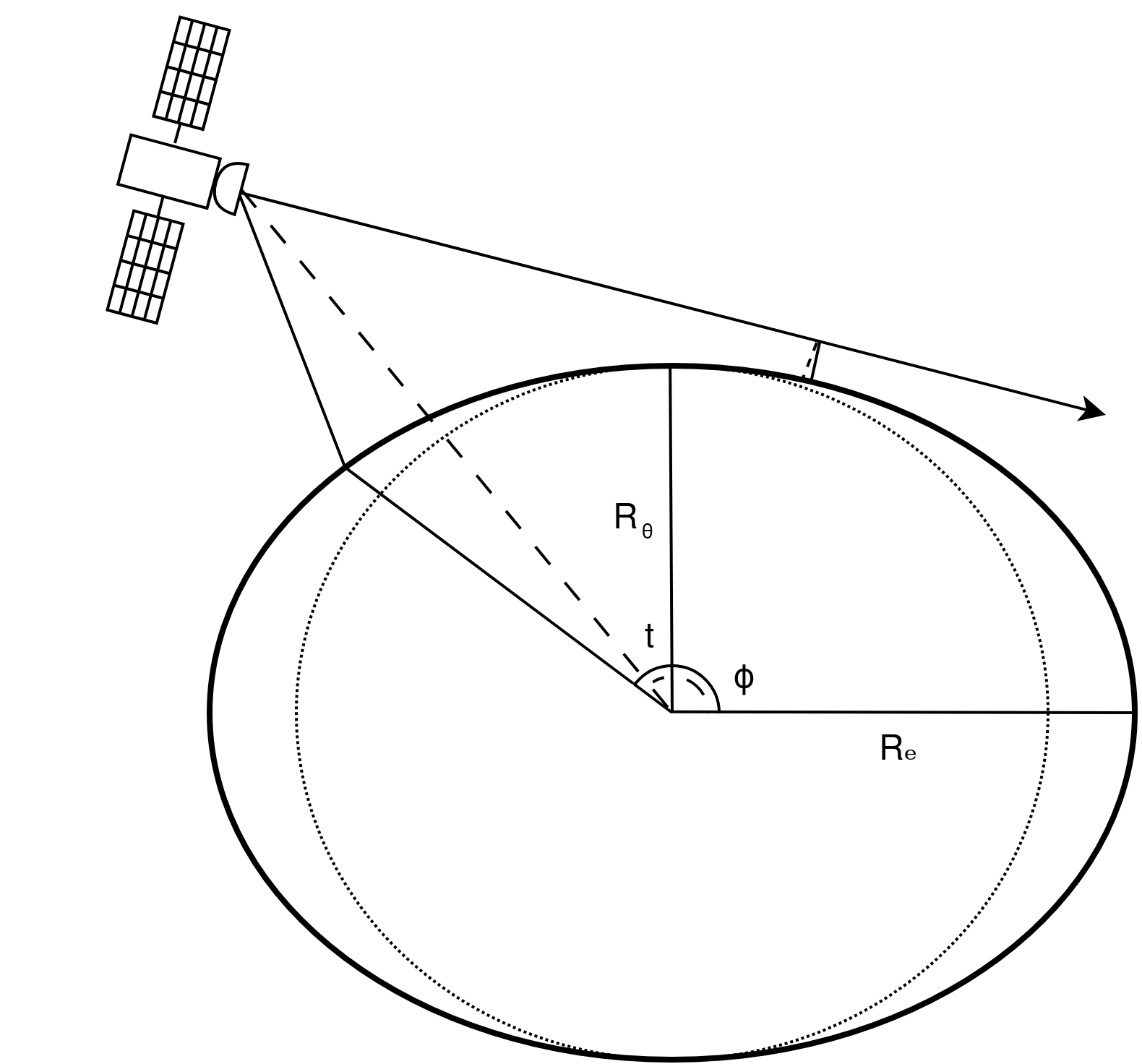}
    \caption{
    The diagram compares viewing geometries for a spherical Earth and an ellipsoidal Earth. In the spherical case, the local vertical (normal to the surface) coincides with the radius of the circle. In contrast, for an ellipsoidal Earth, the local vertical deviates from the radial direction, resulting in a shifted tangent point. Consequently, the tangent altitude also differs between the two cases, illustrating how Earth's shape affects the line-of-sight geometry and the definition of vertical reference.
}
    \label{fig:elliptic_earth}
\end{figure}
%Considering the position of the satellite along its orbit, 
The altitude of a point above the Earth's surface can be determined by tracing the line that minimizes the distance to the surface, which is therefore normal to it. However, in the case of an ellipsoidal Earth, the radial direction defined by the polar coordinate $\phi$ in the orbital plane generally does not align with the surface normal, which defines the surface coordinate $t$, as illustrated in Fig.~\ref{fig:elliptic_earth}. Alignment occurs only at specific angles that are multiples of $90^\circ$, corresponding to points where the orbital plane intersects the equator and the locations of maximum absolute latitude. This misalignment causes a shift in the location of the tangent point and prevents the derivation of a closed-form expression for directly converting the angular coordinate in the orbital plane into a surface coordinate. Several methods for this transformation have been proposed \cite{pewsey_minimum_distance_ellipse,nurnberg_distance_to_ellipse}. On the whole, we found that Chatfield's method \cite{chatfield_distance_to_ellipse}, which uses the ellipse evolute, performs well, and we adopted it.

Once the surface coordinate $t$ is determined, the corresponding altitude $z$ can be computed as the distance from the satellite position to the associated point on the Earth's surface. This defines a coordinate transformation:
\begin{equation}\label{eq:xy_to_tz}
\begin{pmatrix}t \\ z\end{pmatrix}= \boldsymbol{\mathrm{F}}
\begin{pmatrix}x \\ y\end{pmatrix}
\end{equation}
which maps the Cartesian coordinates $(x, y)$ to the curvilinear coordinates $(t, z)$. This transformation is used in the computation of the lines of sight so that, for our purposes, is applied only outside the ellipse representing the Earth's surface.

As for the atmospheric data, they are extracted based on the computed sub-satellite positions (latitude, longitude, UTC) and then resampled into a $(t,\,z)$ grid to build the model.

\subsection{Lines of sight and refractivity}
A line of sight (LoS) is the trajectory followed by an electromagnetic ray. Its shape depends on the speed of light in the medium through which the ray travels. In a vacuum, the speed of light is constant, and the path is a straight line. However, in the atmosphere, the speed of light depends on the refractive index $n$, given by $v = c/n$, where $c$ is the speed of light in a vacuum. Since $v$ is typically close to $c$ in the atmosphere, $n$ is slightly greater than 1. For convenience, this property is often quantified using the refractivity, defined as $N = (n - 1)\times 10^6$, instead of $n$.

The propagation of an electromagnetic ray path in a medium is described by the Eikonal equation \cite{born1999}. It is derived from the wave equation under the assumption that the refractive index $n$ varies slowly compared to the wavelength of the radiation. The LoSs are then the solution of this equation, that can be written \cite{born1999}:
\begin{equation}\label{eq:eikonal}
  \frac{d}{ds}\left(n\big(\bfL(s)\big)\frac{d \bfL(s) }{ds}\right)=\nabla n\big(\bfL(s)\big),
\end{equation}
where $s$ is the arc parameter, $\bfL(s)$ and $n(s)$ are the LoS and refractive index as a function of $s$.

We restrict our study to two-dimensional ray tracing in the orbit plane in Cartesian coordinates $(x,y)$, more details in Appendix. For the numerical solution of the Eikonal equation, it is necessary to calculate the refractive index $n(x,y)$ and their derivatives $n_x(x,y)$, $n_y(x,y)$ as a function of a generic point $(x,y)$. The refractive index $n$ mainly is mainly governed by air density and can be computed using standard models of atmospheric refraction. The Edlén model \cite{edlen1966}, calculates the refraction index as
\begin{equation}
    n = 1 + c_0 \frac{T_0}{p_0} \cdot\frac{p}{T}.
\end{equation}
In this simple model, the refractivity is proportional to the atmospheric air density, weighted with a standard density at sea level ($p_0= 1013.24$ hPa, $T_0=288.16$ K, $c_0 = 0.000272632$). This old model is still perfectly able to give reliable results. However, Ciddor \cite{ciddor1996} proposed a more accurate method that takes into consideration water vapor content and \ce{CO2} concentration. According to Mathar \cite{mathar2004}, Ciddor method has an accuracy in the refractive index of $10^{-6}$ up to $25$ \unit{\micro m}, an interval that includes the full spectral range of MIPAS and CAIRT. Mathar himself \cite{mathar2007} also proposed a different model for the refractivity, aimed at characterizing ambient air at mountain altitudes where astronomical telescopes are located and using the theory of summation of electromagnetic line transitions. A very interesting discussion on the calculation of the refractive index can be found online at \cite{young_air_refractivity}, that also highlights the relationship between the precision of the refractive model and the precision of the quantity involved in the method. 

In this paper, we adopt the Ciddor method, although the difference between the Edlén and Ciddor methods for the spectral range under consideration has been shown to be negligible \cite{ridolfi2014} with respect to other sources of uncertainty.

For solving the Eikonal equation, we use the Adaptive Eikonal (AEIK) method described in \cite{ridolfi2014}, which employs a multi-step predictor-corrector approach and an adaptive step size based on the curvature of the line of sight. This allows for larger step sizes (up to $1$ \unit{km}) in regions with low atmospheric density, while reducing the step size to as small as $10$ \unit{m} in denser layers. A detailed discussion of the numerical challenges in solving the Eikonal equation is provided in the Appendix.

\subsection{Atmospheric models} \label{sec_atmmod}
For our simulations we use two atmospheric models.
\begin{itemize}
    \item The \textit{US Standard Atmosphere 1976 (US76)} \cite{US1976}: This is a reference standard atmosphere that provides average conditions applicable globally. The US76 atmosphere includes profiles of pressure, temperature and water vapour. The profiles of other gases are not used in this study, as only the refractive index is required.
    \item The \textit{Reanalysis (or ERA5) atmosphere}: This is a realistic atmosphere based on reanalysis data, which we assume to represent the true atmosphere. The meteorological data used in this study were obtained from the ERA5 reanalysis dataset provided by the European Center for Medium-Range Weather Forecasts (ECMWF) \cite{hersbach20}. It includes profiles of pressure, temperature and water vapor, while the \ce{CO2} concentration was sourced from the Copernicus Atmosphere Monitoring Service (CAMS) global Greenhouse Gas Reanalysis Data set (EGG4) \cite{egg2020}. The profiles were extended up to $120$ \unit{km} using the IG2 (Initial Guess 2) climatology \cite{remedios07}. The supporting CASIA project aims to develop a forward model for the CAIRT instrument. Thus, the code itself fills the atmosphere with remaining gases and cloud parameters, while also incorporating surface properties for synergy with the IASI-NG instrument. However, for the purposes of this study, only the quantities relevant to the calculation of the LoS are needed.
\end{itemize}

\subsection{MIPAS and CAIRT instrumental settings}
The MIPAS instrument \cite{raspollini2013,dinelli2021,raspollini2022} was an interferometer onboard the ENVISAT satellite that sounded the atmosphere from 2002 to 2012. Each instrument scan collected $27$ sequential sweeps in the nominal mode during the OR (Optimized Retrieval) period, with tangent heights ranging from $6$ to $60$ \unit{km}. The scans were approximatively spaced by $400$ \unit{km}. The initial retrieval codes did not account for horizontal atmospheric variability. As a result, ray tracing was not necessary; it was sufficient to use Snell's law and an integral transformation to convert line integrals into vertical domain integrals. However, M. Kiefer \cite{kiefer2010} later discovered that neglecting horizontal gradients introduced a small bias in the retrieval of some species, which became evident when performing zonal averages for the same latitude bands in the ascending and descending parts of the orbit. Consequently, the retrieval codes were updated to include the ray tracing. Version 8 of the ESA operational code, the Optimized Retrieval Model (ORM) \cite{raspollini2022}, models the horizontal gradients based on a previous reprocessing, or estimated from reanalysis data. The paper \cite{ridolfi2014} explores the difference in tangent altitudes when using the US76 atmosphere (as used in the level 1 files to produce the engineering tangent altitudes) versus the retrieved atmosphere, with horizontal gradients calculated from the difference between adjacent scans.

The CAIRT instrument is one of the two candidates for the ESA Earth Explorer 11 mission. While the project is still in development, it is expected to include a 2D matrix detector, capable of providing sufficient spectra to enable three-dimensional retrievals. Given the large number of spectra generated, an efficient ray-tracing algorithm will be essential. In this study, we focus on the 2D retrieval within the orbital plane, where the azimuth angle is set to zero.

\section{Results} \label{se3:Results}
\subsection{Determination of the nadir angles}
\begin{figure}[ht!] 
    \centering
    \includegraphics[width=0.55\linewidth]{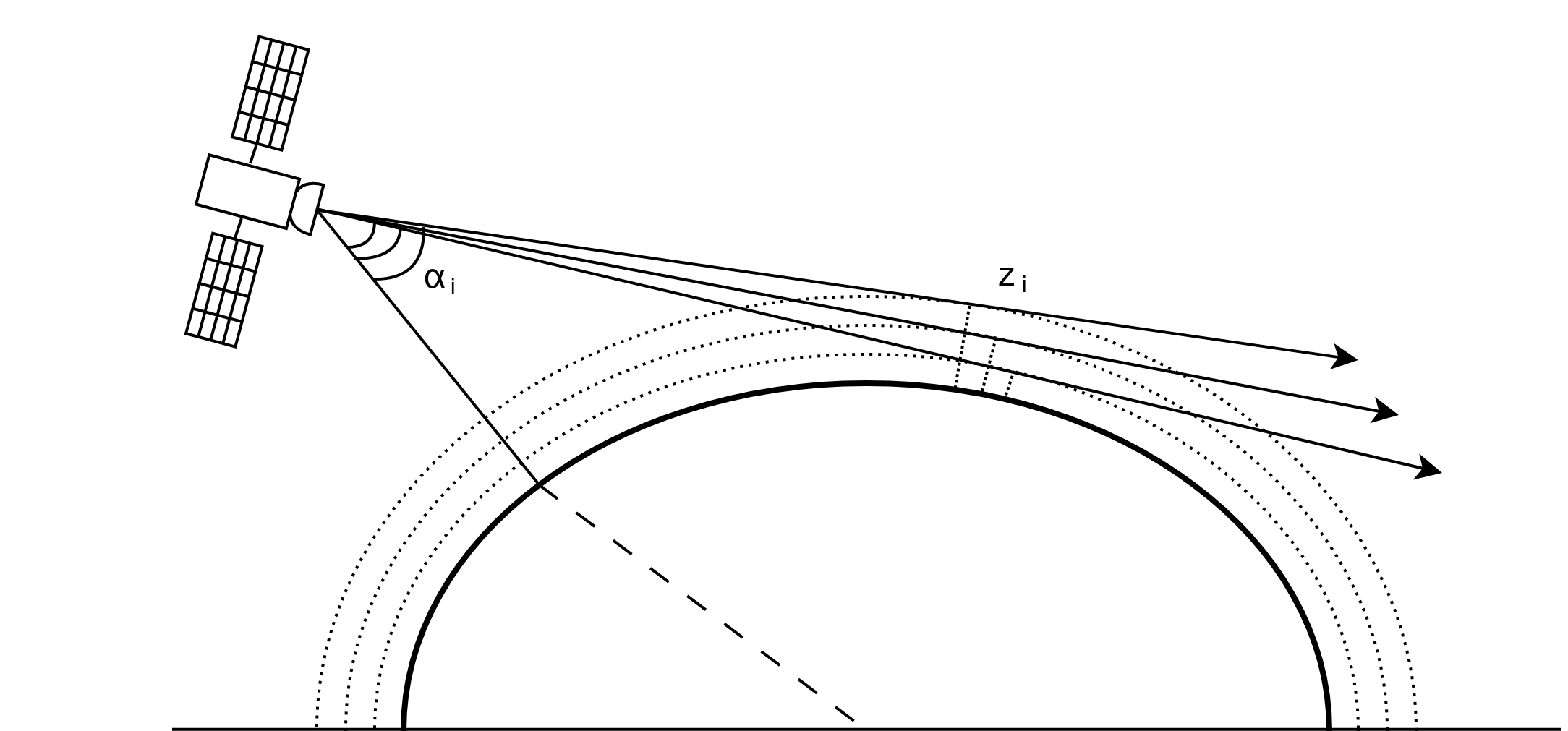}
    \caption{Relationship between nadir angles and tangent altitudes for a geoid-shaped Earth. Nadir angles are measured between the local vertical (normal to the Earth's surface) and the line-of-sight directions. Each nadir angle corresponds to a different tangent altitude, determined by the curvature of the Earth and the geometry of the line of sight.}
    \label{fig:nadir}
\end{figure}

A crucial aspect of limb observation is the determination of the viewing  geometry. In a 2D model, it can be described by the nadir angles. These angles are defined as the angles between the direction of the satellite detectors and the vertical line from the satellite to the Earth's surface. The set of nadir angles $\alpha_i$ should be predetermined so that, in principle, the lines of sight correspond to a predetermined set of tangent point altitudes $\tgz_i$, referred to as the \emph{engineering} tangent altitudes. Although the radiation collected by each detector travels along the line of sight toward the instrument, the ray-tracing algorithm operates in the opposite direction. It begins at the satellite, using the nadir angles to determine the entry points into the atmosphere along a straight line. From there, it follows the solution of the Eikonal equation, descending to the tangent point at the minimum altitude before continuing through the atmosphere until the line of sight exits it. A diagram illustrating the relationship between lines of sight and tangent altitudes is presented in Figure~\ref{fig:nadir}.

Under operational conditions, the relationship between the nadir angles and the tangent points depends primarily on two factors. Both of these factors must be considered to maintain a nearly constant tangent point altitude along the orbit.

The first factor is the satellite's position in the orbital plane, as it determines the sector of the Earth's elliptical shape traversed by the line of sight (LoS) and, consequently, its curvature. This effect can be anticipated and compensated for, since the shape of Earth's section remains fixed as long as the orbital inclination does not change.

The second factor is the atmospheric profile along the LoS. Unlike the geometric effect, this factor can only be mitigated. Unless the atmospheric state is known in advance, it is not possible to fully account for the spatial variation of the refractive index, which governs the curvature of the LoS and, in turn, the altitude and position of the tangent point. The refractive index is directly influenced by the vertical profiles of temperature, pressure, and humidity, which determine air density and, consequently, refractivity. These atmospheric parameters cause fluctuations in the LoS curvature, impacting the location of the tangent point.

Therefore, while the satellite's position and orbital geometry can be accurately predicted and corrected, variability in atmospheric conditions introduces an element of uncertainty that cannot be entirely eliminated without prior knowledge of the atmospheric state. We adopt the following strategy to determine an initial set of nadir angles that compensate for the Earth's ellipticity:

\begin{enumerate}
    \item[(i)] We sample the nadir angle interval $[\an, \ax]$ with a set $\{\alpha_i\}$ such that:
    \begin{equation}
        \alpha_0 = \an, \ \dots, \ \alpha_i < \alpha_{i+1},\ \dots, \ \alpha_M = \ax, \qquad i = 1, \dots, M-1.
    \end{equation}
    This sampling does not need to be uniform and does not depend on the particular scan. Based on geometric considerations, the range of nadir angles that result in a tangent point within the atmosphere may be restricted to the interval $61^\circ = \an < \alpha < \ax = 65^\circ$.

    \item[(ii)] Next, we define a set of polar coordinates of interest. These may correspond to the instrument’s positions $(r\cos(\phi_j), r\sin(\phi_j))$ in the orbital plane at the times of acquisition. Alternatively, they may be a representative sampling of the entire orbit:
    \begin{equation}
        \phi_0 = 0^\circ, \ \dots, \  \phi_j = j\Delta\phi, \ \dots, \ \phi_N = 360^\circ, \qquad j = 0, \dots, N-1.
    \end{equation}

    \item[(iii)] For each coordinate $j$, we trace the lines of sight $\textbf{L}(s)_{ij}$ corresponding to the set $\{\alpha_i\}$, obtaining a new set of tangent altitudes $\{z_{ij}\}$, with $i = 0, \dots, M$ and $j = 0, \dots, N$. We then define a linear spline $s_j(z)$, built using the nodes:
    \begin{equation}
        (z_{ij}, \alpha_i), \quad i = 0, \dots, M, \quad j = 0, \dots, N.
    \end{equation}

    \item[(iv)] Finally, we obtain the required nadir angles $\alpha_{ij} = s_j(\tgz_i)$ by sampling the splines at the engineering tangent altitudes $\{\tgz_i\}$.
\end{enumerate}

To account for atmospheric effects on the ray path, we consider two models
\begin{itemize}
    \item {\bf Geometric model:} The lines of sight are computed using straight lines, neglecting the bending effect of atmospheric refraction. In this case, the nadir angles depend solely on the Earth's shape and can be computed analytically.
    \item {\bf US76 model:} The lines of sight are computed assuming a standard US76 atmosphere, a globally uniform atmospheric model previously introduced. In this case, there is no analytical expression for the ray path, and the Eikonal equation must be solved numerically.
\end{itemize}

\subsection{Accuracy of the tangent point prediction}
We study here the displacement of the tangent point as predicted by one of the two models (geometric and US76), in comparison with the tangent point calculated with the reference reanalysis atmosphere, as described in Section~\ref{sec_atmmod}. While the CAIRT instrument samples up to the mesosphere, the influence of the atmosphere is only relevant for lower altitudes, so we focus on these. The polar coordinate of the satellite in the orbital plane is used to represent the position of the acquisition along the orbit. However, since the sensor looks backwards, there is a shift of about $25^\circ$ between the polar coordinates of the satellite and the tangent point position. This offset must be taken into account when locating the tangent points.

\begin{figure}[t!] 
    \centering
    \includegraphics[width=0.95\linewidth]{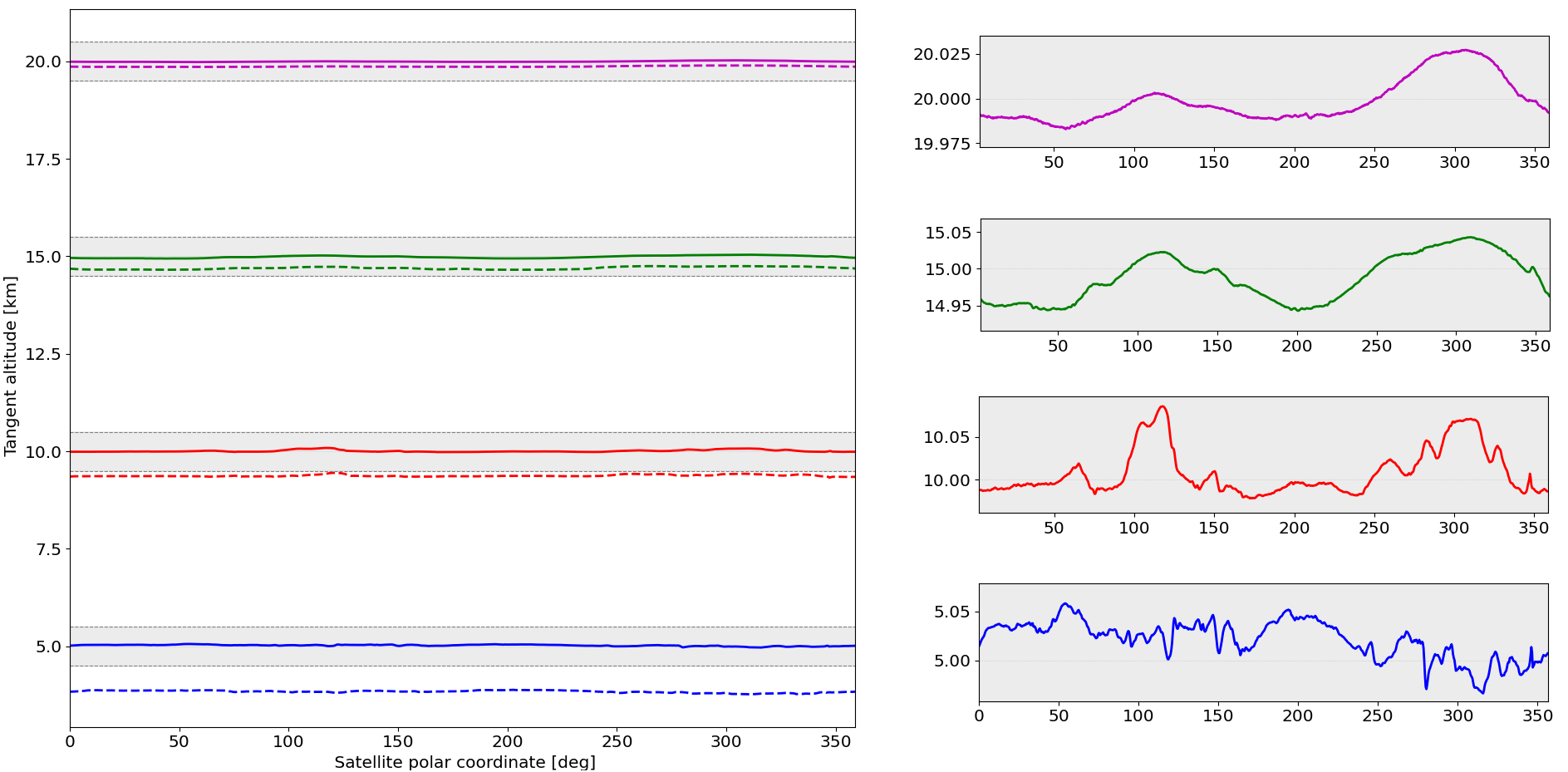}
    \caption{Variation of the tangent altitude as a function of the satellite polar coordinate for a test orbit. Left panel: tangent altitudes obtained with the reference atmosphere along the orbit when the nadir angles are estimated with the geometric model (dashed line) and with the US76 model (solid line). A tolerance of $500$~\si{m} is also shown. Right panel: zoom of the US76 plots}
    \label{fig:fig_3.1}
\end{figure}

We selected a test sun-synchronous orbit with an average altitude of $830$~\si{km}, the ascending node at a latitude of $0^\circ$, and the UTC time of July 10, 2021, 12:00:00. In the left panel of Figure~\ref{fig:fig_3.1}, we show the tangent altitudes along the full orbit for the $5$, $10$, $15$, and $20$ kilometer engineering altitudes, when the associated nadir angles are predicted with the geometric (dashed line) or US76 models (solid lines). A tolerance band of $500$~\si{m} is also shown. This value corresponds to the goal of the \emph{absolute geometric knowledge} as specified in the CAIRT Mission Required Document (ESA, private communication). In the right panel, we show the details of the tangent altitudes predicted using the US76 model.

As expected, the predictions from the geometric model are less accurate than those made with the US76 model. The nadir angles required to reach the engineering tangent altitudes are underestimated in the geometric model, as the curvature effect of the atmosphere is not accounted for. As a consequence, the tangent point altitudes exhibit a negative bias. 

%Smaller nadir angles induce negative biases on the effective tangent altitudes.
\begin{figure}[ht!] 
    \centering
    \includegraphics[width=0.68\linewidth]{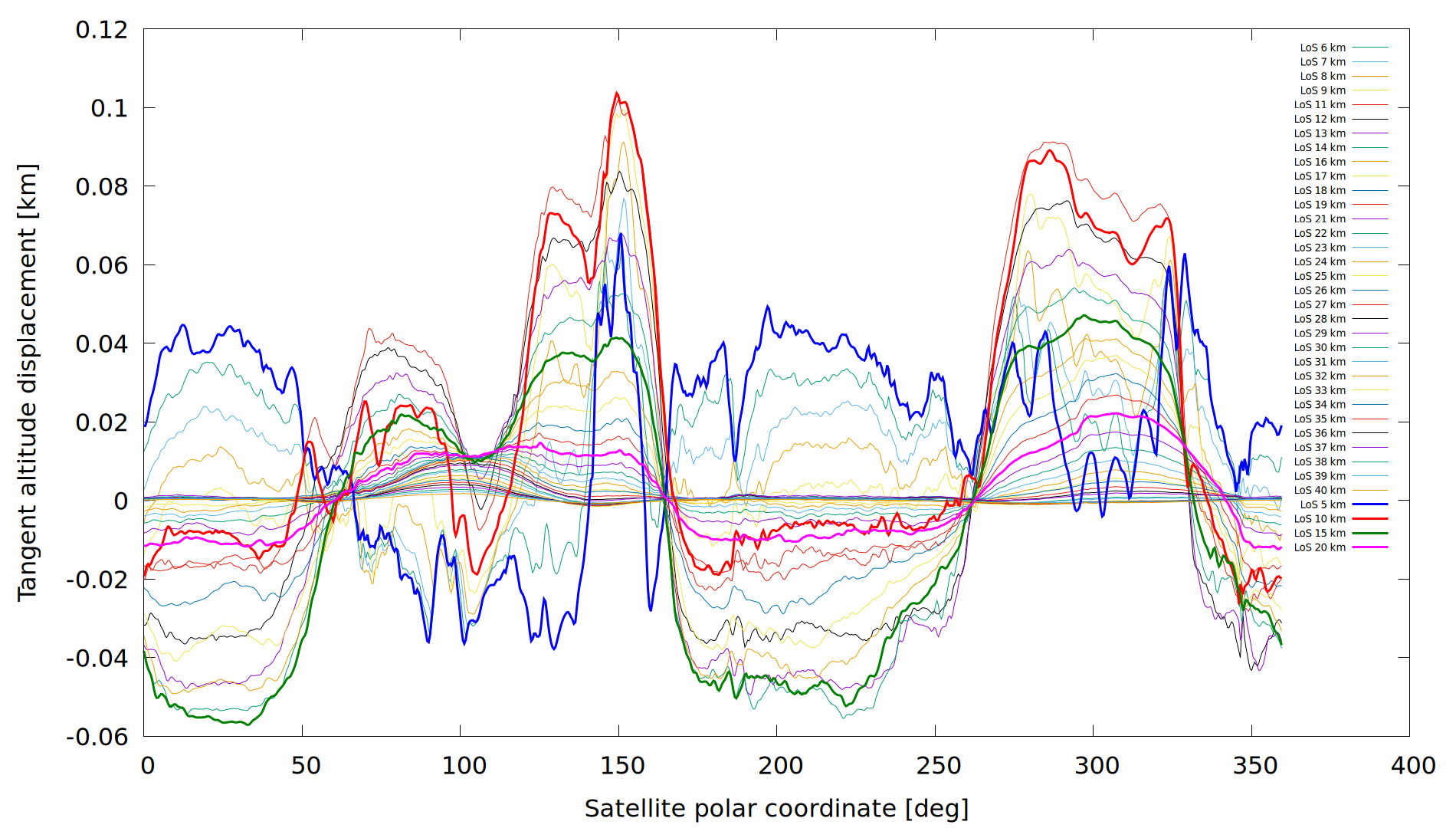}
    \caption{Tangent altitude vertical displacement as a function of the polar coordinate of the satellite. The tangent altitudes considered in the previous figure are highlighted with a thicker line, and the color is maintained}
    \label{fig:fig_los}
\end{figure}
In Figure~\ref{fig:fig_los}, we provide a more detailed view of the displacement of the tangent altitudes for engineering values ranging from $5$ to $40$ \si{km}, with a $1$~\si{km} step for the US76 model. The values presented in Figure~\ref{fig:fig_3.1} are highlighted with a thicker line in the same color. 

As expected, the difference statistically tends to decrease to zero for higher tangent altitudes. However, there are some exceptions of this trend due to local atmospheric conditions. For example, the largest displacement (about $100$~\si{m}) occurs for the $10$~\si{km} curve.

\begin{figure}[ht] 
    \centering
        \includegraphics[width=0.78\linewidth]{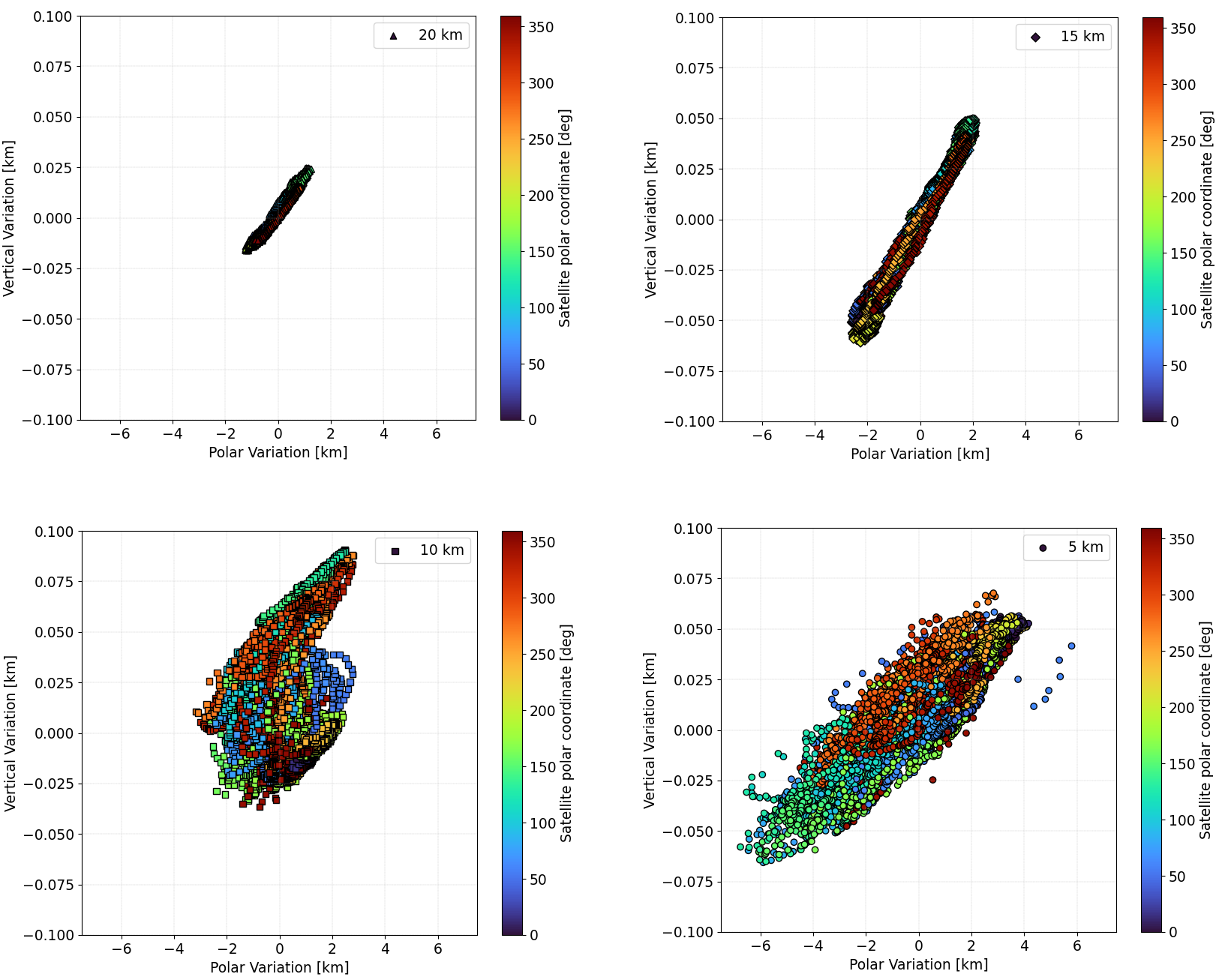}
    \caption{Scatter plots illustrates the variation of the tangent point from nadir angle prediction using the US76 atmosphere model, as a function of the satellite's polar angle. The plots includes data for four reference altitudes: 5 km, 10 km, 15 km, and 20 km, across 14 orbits.}
    \label{fig:displacement}
\end{figure}

The displacement of the tangent points can be measured not only through the vertical difference, as shown so far, but also through the polar coordinate difference, which provides additional spatial information. This can be calculated as the length of the $z$ engineering value level line between the $t$ coordinates of the predicted and calculated tangent points. In the following test, we selected a sample of $14$ consecutive orbits, resulting in different latitude strips that samples the whole Earth. In Figure~\ref{fig:displacement}, we present the results of the test for the four tangent altitudes previously mentioned. The position of the dots represent the polar and vertical variation between the predicted and calculated tangent points. The colormap shows the positions of the points along the orbit, represented by the polar coordinate of the satellite.

We note that there is a correlation between the vertical and the polar displacement. This can be explained because a lower tangent altitude is statistically reached with a longer line of sight, so that the polar coordinate of the tangent point shift backward.

There is also a weaker correlation between the hemisphere being sampled and the displacement of the tangent point. The northern hemisphere is sampled in the first half of the orbit (always considering the angular offset of $25^\circ$), while the southern hemisphere is sampled in the second part of the orbit. This can be explained by the seasonal variability. 
\begin{figure}[ht] 
    \centering
        \includegraphics[width=0.7\linewidth]{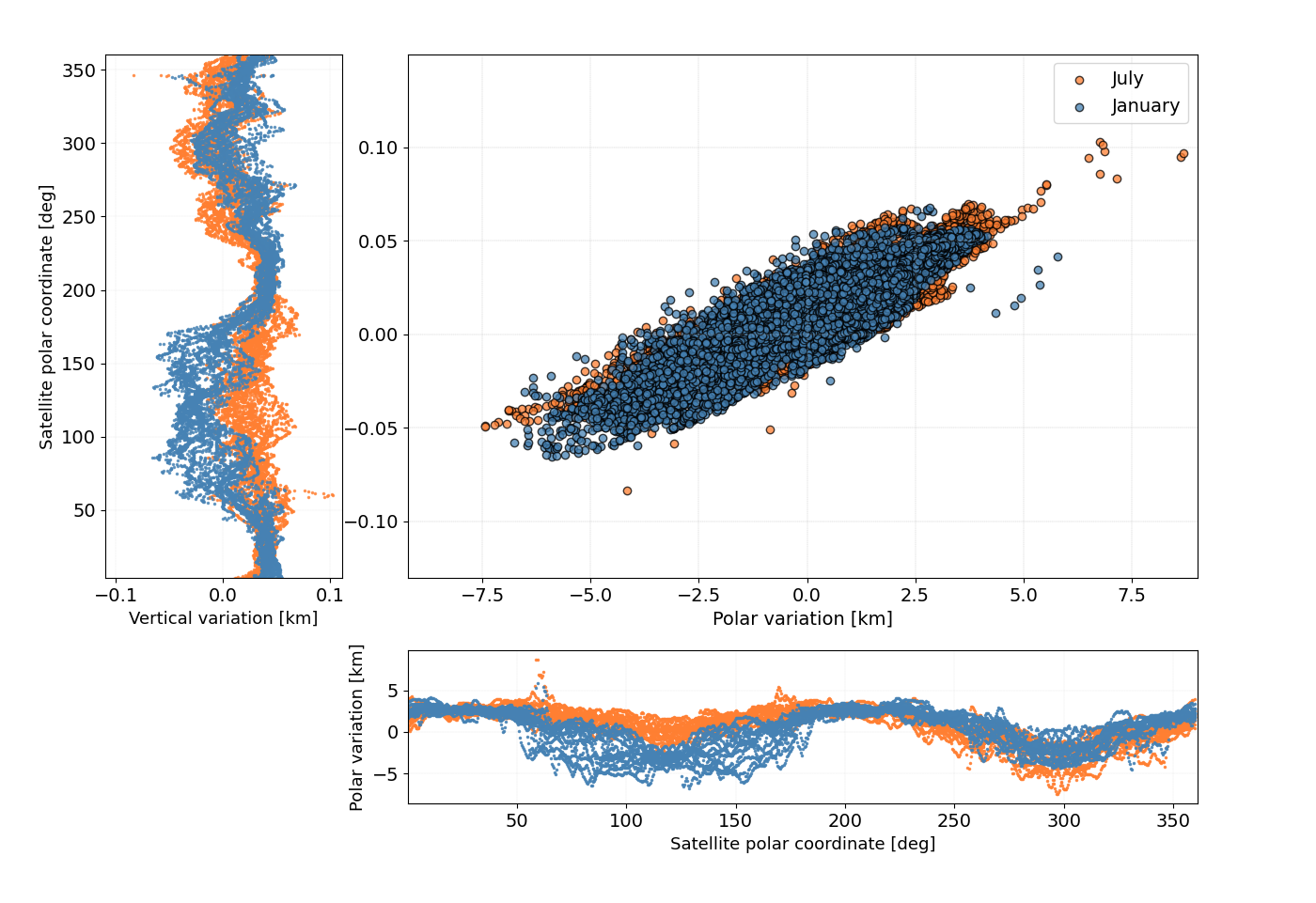}
    \caption{Scatter plot for the $5$~\si{km} tangent altitude. The plot shows the seasonal difference for one orbit in July (orange dots) and one orbit in January (blue dots) compares two distinct atmospheric conditions The differences between the two atmospheric conditions are shown in terms of the variation in the tangent point. The auxiliary scatter plots break the displacement into vertical and polar differences.}
\label{fig:seasons}
\end{figure}

In the test shown in Figure~\ref{fig:seasons}, we compared two orbits covering the same longitudinal band: one acquired in July and the other in January. While the overall scatterplot does not exhibit a distinct pattern, a clear reversal in the displacement trend along the orbit emerges when the displacement is separated into its vertical and polar components. This reversal reflects the transition between different climatological zones. Notably, as the tangent points approach the equator, corresponding to satellite positions around $25^\circ$ and $205^\circ$, the seasonal variation tends to zero.

\section{Comparison between MIPAS and CAIRT ray-tracing} \label{se4:MIPAS-CAIRT}
A direct comparison between the ORM ray-tracing algorithm and the new CAIRT algorithm is not straightforward, due to fundamental differences in their design.

The ORM code assumes fixed polar coordinates for the tangent points, initially determined by a level-1 ray-tracing algorithm based on the US76 reference atmosphere. In the first phase of retrieval, the tangent altitudes are updated by retrieving the corresponding pressures and temperatures, allowing the use of hydrostatic equilibrium to refine the tangent altitudes. This updated information is incorporated into the retrieval grid at each iteration. Ray tracing in ORM begins at the tangent point and proceeds bidirectionally up to the top of the atmosphere. As each scan is processed independently, the Earth is modeled as a sphere, with a local radius computed from the scan’s geolocation. In the most recent version of ORM, horizontal gradients are included, either derived from a prior retrieval that neglects gradients or obtained from ERA5 reanalysis data. ORM applies a numerical approach to approximate derivatives (see the Appendix), which we show may introduce a bias. However, because the algorithm starts at the tangent point, error accumulation is minimized, and the issue has largely gone unnoticed. Finally, it is worth noting that the MIPAS lowest tangent altitude varies with latitude to follow the tropopause, reducing the likelihood of cloudy scenes, so that low tangent altitudes of $6$~\si{km} are attained only near the equator.

In contrast, the CAIRT ray-tracing algorithm operates on a two-dimensional grid defined in altitude and polar coordinates, which can be refined as needed. For our comparison, we adopt a polar step size of $0.45^\circ$, corresponding to approximately $50$~\si{km} between adjacent tangent points, comparable to the expected spacing between CAIRT scans. Our objective is to assess the impact of atmospheric discretization—specifically, the horizontal resolution difference between MIPAS and CAIRT—on the accuracy of tangent altitude determination.

We employed a scanning pattern with angular steps of $3.6^\circ$, closely matching the horizontal separation between successive MIPAS scans (approximately $400$~\si{km}). We then conducted ray tracing using an atmospheric model sampled at this coarser resolution and compared the results to those obtained using the same scanning geometry but with the atmosphere sampled at the finer CAIRT resolution of $0.45^\circ$. In this setup, the CAIRT configuration preserves finer-scale horizontal atmospheric features, whereas the MIPAS configuration effectively smooths out oscillations below the $3.6^\circ$ scale.

\begin{figure}[t!]
    \centering
        \includegraphics[width=0.78\linewidth]{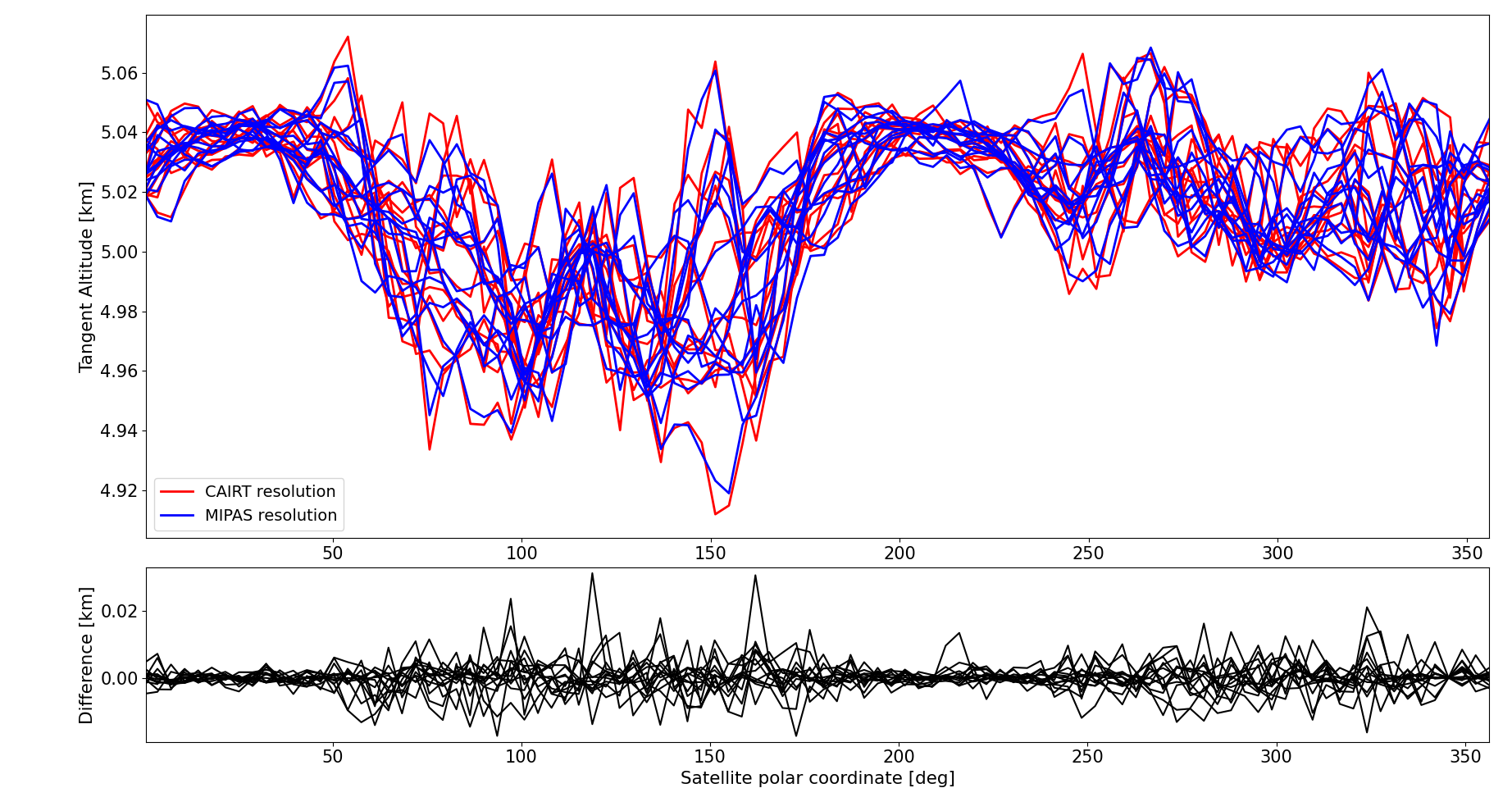}
    \caption{The plot illustrates the variation of the tangent point altitude as a function of the satellite's polar coordinate for a reference tangent altitude of 5~\si{km}. The blue curve represents values computed using the MIPAS horizontal resolution, while the red curve corresponds to those obtained using the finer CAIRT resolution. Tangent heights were evaluated across 14 consecutive satellite orbits. The subplot displays the residuals, defined as the differences between the tangent altitudes estimates derived from the two spatial resolutions.
} \label{fig:mipas-cairt}
\end{figure}

Our analysis focuses once more on the lower tangent altitude of $5$~\si{km}, where atmospheric density is highest and ray tracing is therefore more sensitive to local variations in the refractive index, primarily driven by changes in atmospheric density.

To ensure global coverage, we again examined the set of 14 consecutive orbits already introduced. The results are presented in Figure~\ref{fig:mipas-cairt}. Tangent altitudes computed using the atmosphere sampled at CAIRT resolution are shown in red, while those using the coarser MIPAS resolution are plotted in blue. The lower panel shows the differences between the two. We observe that the smoothing error introduced by the MIPAS resolution remains globally below $20$~\si{m}. This represents an additional uncertainty that cannot be captured using MIPAS data alone due to its limited horizontal resolution.

\begin{figure}[ht]
    \centering
        \includegraphics[width=1\linewidth]{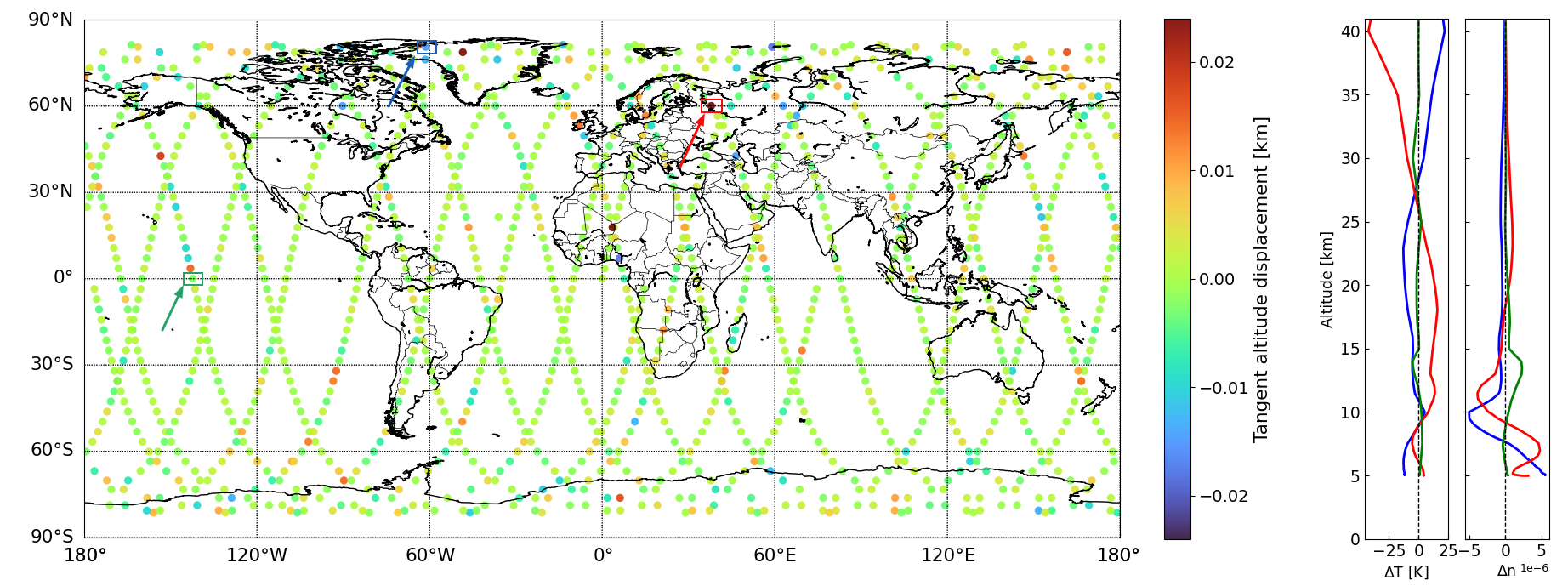}
    \caption{Geolocated tangent point differences between CAIRT and MIPAS resolutions. The dots color indicates the magnitude and sign of the difference. Three representative cases are highlighted: one with negligible difference (green), one with a large positive difference (red), and one with a large negative difference (blue). The side panels show the corresponding temperature and refractive index displacement profiles along the line of sight for each of the three selected cases.} \label{fig:mc_earth}
\end{figure}
To further investigate the origin of the observed differences, we analyzed their spatial distribution across the globe. In Figure~\ref{fig:mc_earth}, we present a global map of the differences in tangent point altitudes between the MIPAS and CAIRT resolutions. Three representative points are highlighted with arrows: one showing a negligible difference (green), and two showing significant positive (red) and negative (blue) differences. In the right panels, we display the corresponding refractive index and temperature profile differences along the lines of sight for these selected points. As expected, the refractive index and temperature are nearly identical for the point with negligible difference. In contrast, for the other two cases, the sign and magnitude of the differences correlate with the sign of the tangent altitude discrepancies. These patterns reflect how the distinct horizontal resolutions of the MIPAS and CAIRT configurations respond differently to atmospheric variability along the line of sight.

The global discrepancy between the US76 standard atmosphere and the reference atmosphere was previously assessed to be below $200$~\si{m} for the MIPAS instrument~\cite{ridolfi2014}. In this work, we found that the corresponding discrepancy for the CAIRT instrument is below $100$~\si{m}. While the discrepancies are of the same order of magnitude, and acknowledging the differences in algorithmic approaches, we observe a reduction in bias for CAIRT. This improvement can be attributed to both the finer horizontal resolution of the CAIRT instrument and the more enhanced ray-tracing algorithm.

\section{Conclusions}
In this paper, we studied a ray-tracing algorithm that can be applied to the CAIRT instrument. Limb-scanning interferometers have a vertical resolution much better than nadir scanning sensors, but they have to deal with ray tracing, to compensate for the refraction due to the atmosphere. The atmospheric variability influences the tangent point position. However, the pointing angles must be set before the actual atmospheric condition traversed by the radiation reaching the instrument's detectors is known. Thus, an estimate based on a constant standard atmosphere could be used. We showed that this uncertainty can reach $100$~\si{m} in altitude, and several kilometers along the polar coordinate.

We then analyzed the dependence of the bias on the various factors, such as the position of the tangent point along the orbit, the atmospheric variability or the seasonal effect.

Finally, we assessed the smoothing error due to lower resolution of the MIPAS instrument compared to the new CAIRT instrument. This error could not be assessed from the MIPAS data, because the horizontal atmosphere could only be discretized using the MIPAS resolution itself. We verified that the error is at most $20$~\si{m}, a value that is lower than the accuracy obtained for the MIPAS data. We observed a reduction of the discrepancy in the CAIRT tangent altitudes with respect to those estimated for MIPAS. This is due both to the better horizontal resolution of the CAIRT instrument, and to the improvements of the ray-tracing numerical algorithm presented in the Appendix.

Finally, some considerations about the computation time. Once the radius and inclination of the orbit is known, the corresponding predicted nadir angles can be determined all along the entire orbit. This step is not required at every iteration, as the predicted nadir angles can be precomputed once and reused for all subsequent calculations.

The time required to compute a single LoS depends primarily on the length of its intersection with the atmosphere. Lower tangent altitudes result in longer atmospheric paths and consequently higher computational costs. For example, LoS paths with tangent altitudes around $5$~\si{km} are significantly more demanding than those at higher altitudes. Although the computation time per LoS remains moderate on standard hardware, the large number of LoS paths required by CAIRT may lead to a computational bottleneck. As a reference, on a standard Intel i7 processor, the ray-tracing code requires approximately 2 hours to process a full orbit consisting of 800 scans, each containing 85 LoS.

To mitigate this issue, we foresee two possible optimization strategies:

\begin{itemize}
    \item[(i)] \textbf{Parallelism:} The calculation of each line of sight is independent of the others, allowing for straightforward parallelization of the code without requiring any modification to the underlying algorithm.
    
    \item[(ii)] \textbf{AI methods:} Given the observed correlations between tangent altitude differences and factors such as orbital position and seasonal atmospheric variability, it may be possible to improve tangent altitude predictions using machine learning models trained on a representative dataset. This approach could enable more accurate estimation of the required nadir angles, thereby reducing the displacement between predicted and actual tangent altitudes.
\end{itemize}

An additional optimization could involve reducing the number of lines of sight explicitly computed by the ray-tracing algorithm. Lines of sight corresponding to similar engineering tangent altitudes are often strongly correlated, as they traverse nearly identical portions of the atmosphere. Therefore, it may be feasible to interpolate or approximate intermediate lines of sight using machine learning techniques trained on a reduced set of fully computed paths.

However, a key challenge arises from the level of detail required by the forward model: the relevant information is not limited to scalar quantities, but includes the precise geometry and trajectory of each ray path. This geometric fidelity must be preserved to ensure the accuracy of the radiative transfer calculations, making interpolation or emulation non-trivial.

\section*{Declarations:}
\subsection*{Funding}
The work presented was supported by the CASIA project of the ASI (Agenzia Spaziale Italiana),\break CUP: F93C23000430001.
\subsection*{Conflict of interest}
The authors declare no conflict of interest.
\subsection*{Data availability}
The atmospheric and surface scenarios and numerical results of the tests presented in this paper are available at \url{https://doi.org/10.5281/zenodo.15393518}. 
\section*{CRediT authorship contribution statement}
\noindent{\bf Luca Sgheri:} Methodology, Formal Analysis, Software, Writing – original draft, Writing – review \& editing, Funding Acquisition;

\noindent{\bf Francesco Pio De Cosmo:} Formal Analysis, Software, Testing, Writing – original draft, Writing – review \& editing;
\appendix
\section{Numerical resolution of the Eikonal equation}
The Eikonal equation (\ref{eq:eikonal}) can be solved in Cartesian coordinates by setting $\bfL(s)=(x(s),y(s))$. It can be written as a system of second order differential equations: 
\begin{equation}\label{eq:eik}
  \left\{\begin{aligned}
    n^\prime \big(x(s),y(s)\big) x^\prime(s) +  n\big(x(s),y(s)\big) x^{\prime\prime}(s) = n_x\big (x(s),y(s)\big) \\
    n^\prime\big(x(s),y(s)\big) y^\prime(s) +  n\big(x(s),y(s)\big) y^{\prime\prime}(s) = n_y\big(x(s),y(s)\big)
  \end{aligned}\right.
\end{equation}
For the numerical implementation the system can be rewritten using auxiliary variables $y_i$:
\begin{equation}\label{eq:eik_cart1}
  \left\{\begin{aligned}
  y_1(s)&=x(s)\\
  y_2(s)&=y(s)\\
  y_3(s)&=n\big(x(s),y(s)\big)x'(s)\\
  y_4(s)&=n\big(x(s),y(s)\big)y'(s)
  \end{aligned}\right. \qquad \qquad \qquad  \left\{\begin{aligned}
  y'_1(s)&=\frac{y_3(s)}{n\big(y_1(s),y_2(s)\big)}\\
  y'_2(s)&=\frac{y_4(s)}{n\big(y_1(s),y_2(s)\big)}\\
  y'_3(s)&=n_x\big(y_1(s),y_2(s)\big)\\
  y'_4(s)&=n_y\big(y_1(s),y_2(s)\big)\\
  \end{aligned}\right.
\end{equation}
Combining the system with initial conditions $y_i(0) = y_i^0$, the equation can be solved to determine the line of sight (LoS). Although the satellite position and viewing angle could be used as initial conditions, a more practical approach is to initialize the integration at the point where the LoS, extending from the instrument with the given inclination, intersects the uppermost atmospheric level. Since the LoS is a straight line outside the atmosphere, this intersection point can be determined analytically even when the Earth is modeled as a geoid.

We employ the Adaptive Eikonal (AEIK) method \cite{ridolfi2014}, a three-level predictor-corrector scheme that offers a good balance between computational efficiency and accuracy. This method shows a slight performance advantage \cite{ridolfi2014} compared to the ray-tracing algorithm implemented in the \textit{KOmmunikations und PRojekt Analyse} (KOPRA) code \cite{kopra}, a widely recognized reference in the field.

We define $\mathbf{y} = (y_1, \dots, y_4)^\top$ and $\mathbf{f} = (y'_1, \dots, y'_4)^\top$, see Equation~(\ref{eq:eik_cart1}). For the predictor step, we employ the Adams–Bashforth formula:
\begin{equation}
\mathbf{y}_{n+2}^{(p)} = \mathbf{y}_{n+1} + \frac{3}{2} \Delta s\, \mathbf{f}(s_{n+1}, \mathbf{y}_{n+1}) - \frac{1}{2} \Delta s\, \mathbf{f}(s_n, \mathbf{y}_n),
\end{equation}
while the corrector step uses the second-order Backward Differentiation Formula (BDF2). Although BDF2 is typically an implicit multistep method used to solve stiff ordinary differential equations, here it is applied in an explicit form:
\begin{equation}
\mathbf{y}_{n+2} = \frac{4}{3} \mathbf{y}_{n+1} - \frac{1}{3} \mathbf{y}_{n} + \frac{2}{3} \Delta s\, \mathbf{f}(s_{n+2}, \mathbf{y}_{n+2}^{(p)}).
\end{equation}
The three-level scheme is initialized using a combination of the explicit and implicit Euler methods. This numerical integration approach is also adopted in the ORM code.

Ray tracing is the first step of the forward model for a limb-scanning instrument, defining the curved LoS paths through the atmosphere. To compute radiative transfer, the optical depth must be integrated along each path segment intersected by the LoS. Following MIPAS terminology, a \emph{clove} refers to a sector of an elliptical annulus in the atmospheric grid, and a \emph{path} is the intersection of a LoS with a clove.

For the forward model, Curtis-Godson (CG) integrals \cite{godson1953} of various atmospheric quantities must be computed along each path. The CG integral $\bar{c}$ of a generic quantity $c$ is defined as:
\begin{equation}
\bar{c} = \frac{\int_{s_i}^{s_{i+1}} c\big(\mathbf{L}(s)\big)\, \rho\big(\mathbf{L}(s)\big) \, ds}{\int_{s_i}^{s_{i+1}} \rho\big(\mathbf{L}(s)\big) \, ds},
\end{equation}
where $[s_i, s_{i+1}]$ denotes the arc length segment of the path within clove $i$, and $\rho$ is the air density, which is proportional to $p/T$. The integral is approximated using a quadrature formula that combines internal points collected during the Eikonal equation solution with two additional boundary segments. These boundary segments may significantly contribute to the integral when $\Delta s$ is large.

To accelerate the quadrature evaluation, we assume that the step size $\Delta s$ remains constant within each clove and only changes at clove boundaries, so that the internal quadrature nodes are evenly spaced. At each boundary crossing, specifically at step $n+2$, we evaluate the quantity  
\begin{equation}
d_2 = \frac{\|\bff(s_{n+2}, \bfy_{n+2}) - \bff(s_{n+1}, \bfy_{n+1})\|_2}{\Delta s},
\end{equation}
which serves as a proxy for the local curvature of the LoS. Based on the value of $d_2$, we adaptively update $\Delta s$ using a lookup table: as $d_2$ increases, indicating stronger curvature, $\Delta s$ is reduced from $1$~km to $10$~m to maintain accuracy.

The lookup table is calibrated so that, using an upper-bound estimate of the quadrature error, each CG integral is evaluated with an error below $1\%$. Maintaining a conservative integration step is crucial for ensuring this level of precision.

A commonly used strategy to improve accuracy, doubling the number of quadrature nodes, has no effect in our case unless the LoS itself is recomputed. This is because simply inserting additional points into the piecewise-linear LoS and evaluating the integrand via bilinear interpolation (except for pressure) does not alter the integral result.

Although these considerations do not directly affect the calculation of the LoS, they impose certain constraints on the ray-tracing algorithm. Specifically, a constant step size $\Delta s$ must be maintained within each clove. This requirement can be satisfied by using the arc length parameter as the independent variable in the solution of the Eikonal equation. Additionally, a conservative adaptive strategy for $\Delta s$ should be employed to ensure numerical stability and accuracy across varying atmospheric conditions.

The key point in the evaluation of the function $\bff$, necessary for the solution of the Eikonal equation, is the calculation of the derivatives $n_x(x(s),y(y))$, $n_y(x(s),y(s))$ for a given arc parameter $s$, which can only be calculated numerically. For simplicity, we now drop the dependence upon $s$. The derivatives can be calculated directly, using the incremental ratios:
\begin{equation}\label{eq:der1}
\begin{cases}
    \displaystyle \frac{dn}{dx}(x,y) \approx \frac{n(x+\Delta x,y) - n(x,y)}{\Delta x}, \\[8pt]
    \displaystyle \frac{dn}{dy}(x,y) \approx \frac{n(x,y+\Delta y) - n(x,y)}{\Delta y}.
\end{cases}
\end{equation}
Note that the refractive index is given as a function of the $(t,z)$ coordinates, so that the transformation $\boldsymbol{\mathrm{F}}$ of (\ref{eq:xy_to_tz}) is anyway required in the computation above.

An alternative method is to use the chain rule, obtaining:
\begin{equation}\label{eq:chain_rule}
\begin{cases}
    \displaystyle \frac{dn}{dx}(t,z) =  \frac{dn}{dt}  \frac{dt}{dx} +  \frac{dn}{dz}  \frac{dz}{dx}, \\[8pt]
    \displaystyle \frac{dn}{dy}(t,z) =  \frac{dn}{dt}  \frac{dt}{dy} +  \frac{dn}{dz}  \frac{dz}{dy}.
\end{cases}
\end{equation}
The inverse of the $\boldsymbol{\mathrm{F}}$ transformation can be represented in closed form. The point $(x,y)$ is defined in the $(t,z)$ coordinates by the analytical formula:
\begin{equation}
\begin{cases}
    \displaystyle x = \cos(t) \left( R_e + z \frac{R(t)}{R_e} \right), \\[8pt]
    \displaystyle y = \sin(t) \left( R_\theta + z \frac{R(t)}{R_\theta}  \right).
\end{cases}
\end{equation}
where $(x,y)$ represents a point at an altitude $z$ measured along the normal to the ellipse at the point $(R_e \cos(t), R_\theta \sin(t))$. The function $R(t)$ is the radius of curvature of the ellipse and is defined by the following formula:
\begin{equation}
    R(t) = \frac{R_e R_\theta}{\sqrt{R^2_e \sin^2(t) + R^2_\theta \cos^2(t)}}.
\end{equation}
The derivatives of Eq. \eqref{eq:chain_rule} can be computed from the inverse Jacobian matrix of the transformation from $(t, z)$ to $(x, y)$. After extensive calculations, we obtain

\begin{equation}
\begin{cases}
    \displaystyle \frac{dz}{dy} = \frac{R_\theta R_e^2 \sin(t)-R_\theta z  [R^\prime(t) \cos(t) - R(t) \sin(t)]}{R(t)\big[R_e^2 \sin^2(t) + R_\theta^2 \cos^2(t) + z R(t)\big]}, \\[8pt]
    \displaystyle \frac{dt}{dy} = \frac{R_\theta \cos(t)}{R_e^2 \sin^2(t) + R_\theta^2 \cos^2(t) + z R(t)}
\end{cases}
\end{equation} 
\begin{equation}
\begin{cases}
    \displaystyle \frac{dz}{dx} = \frac{R_eR_\theta^2 \cos(t)+R_e z  [R^\prime(t) \sin(t) + R(t) \cos(t)]}{R(t)\big[R_e^2 \sin^2(t) + R_\theta^2 \cos^2(t) + z R(t)\big]}, \\[8pt]
    \displaystyle \frac{dt}{dx} = - \frac{R_e \sin(t)}{R_e^2 \sin^2(t) + R_\theta^2 \cos^2(t) + z R(t)}
\end{cases}
\end{equation}
where $R'(t)$ indicates the first derivatives of the curvature. 
%\begin{equation}
%    R'(t) = -\frac{R_e R_\theta  (  R_e^2 - R_\theta^2) \sin(t) \cos(t) }{[R^2_e \sin^2(t) + R^2_\theta \cos^2(t)]^{3/2}}.
%\end{equation}
The derivatives $dn/dz$ and $dn/dt$ in \eqref{eq:chain_rule} must be calculated numerically, and depend upon the model for the refractive index adopted, though the primary source of variation for the refractive index is anyway the density of the atmosphere. In the case of a horizontally homogeneous atmosphere such as the US76, $dn/dt\equiv 0$, so that  we only need $dn/dz$.

\begin{figure}[ht] 
    \centering
    \begin{minipage}{0.49\textwidth}
        \centering
        \includegraphics[width=\linewidth]{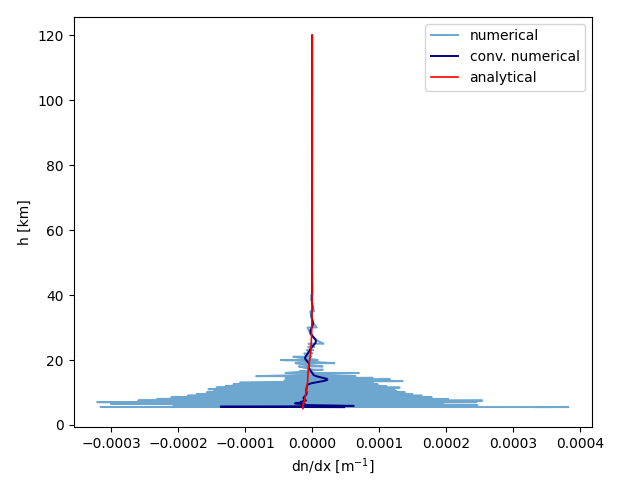}
    \end{minipage}
    \begin{minipage}{0.49\textwidth}
        \centering
        \includegraphics[width=\linewidth]{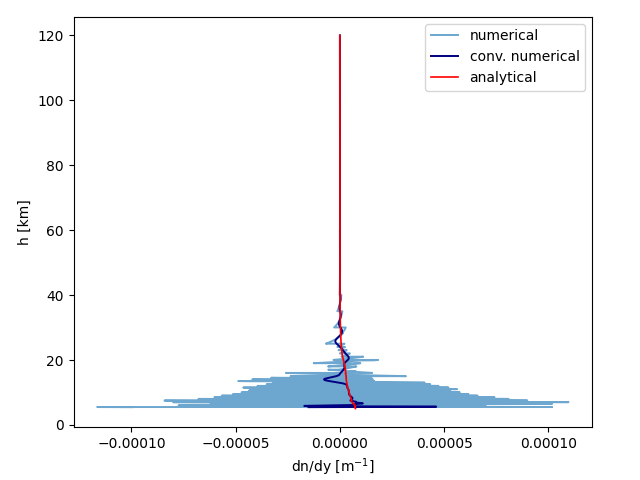}
    \end{minipage}
    \caption{Variation of the refractive index derivatives along x (left) and y (right) with height in the numerically and analytically case.}
    \label{fig:A1}
\end{figure}

From the numerical point of view, the largest variation in density (and hence in the refractive index) is concentrated in the $z$-direction, which is the main reason why  \eqref{eq:chain_rule} is numerically more stable than \eqref{eq:der1}. Also, since $dn/dz$ and $dn/dt$ are normally not differentiable at the clover boundaries, to calculate the derivative we use a three-points local least-squares differentiation formula \cite{savitzky1964}, which produces smoother results.
%Also, the way the derivatives $dn/dz$ and $dn/dt$ are numerically calculated influence the results. The $n$ mainly depends on the density, so has an exponential shape with respect to the altitude. Thus, the analytic derivative of a parabolic interpolation based on three points is preferable even it this means adding a further refraction index calculation. On the other hand, the derivative $dn/dt$ is much smaller and varies linearly, so that a simple forward difference is enough to guarantee a stable result.

Figure \ref{fig:A1} summarizes the results. We plot the $dn/dx$ (left panel) and $dn/dy$ (right panel) along a single line of sight, stopping at the tangent point, as a function of the altitude, and using the ERA5 atmosphere. The light blue curves represent the derivatives calculated directly from \eqref{eq:der1}, the dark blue is the same but convolved with a Gaussian to reduce the impact of numerical fluctuations, while the red curves represent the derivatives calculated using the chain rule \eqref{eq:chain_rule}. We note that the direct calculation suffers from numerical instability, while the red curves are smooth. Actually, the difference in the tangent altitude is not so relevant, because, as it can be seen from the figure, the convolved version retains the shape of the red curve. Statistically, the error in the tangent altitude is lower than $100$~\si{m}, but we observed some scans where the difference is larger than $1$~\si{km}.
\newpage
\bibliographystyle{model1b-num-names}
\bibliography{references}

\begin{thebibliography}{32}
\expandafter\ifx\csname natexlab\endcsname\relax\def\natexlab#1{#1}\fi
\providecommand{\bibinfo}[2]{#2}
\ifx\xfnm\relax \def\xfnm[#1]{\unskip,\space#1}\fi
%Type = Article
\bibitem[{Burrows et~al.(1995)Burrows, Hölzle, Goede, Visser, and Fricke}]{burrows1995}
\bibinfo{author}{J.~Burrows}, \bibinfo{author}{E.~Hölzle}, \bibinfo{author}{A.~Goede}, \bibinfo{author}{H.~Visser}, \bibinfo{author}{W.~Fricke},
\newblock \bibinfo{title}{Sciamachy—scanning imaging absorption spectrometer for atmospheric chartography},
\newblock \bibinfo{journal}{Acta Astronautica} \bibinfo{volume}{35} (\bibinfo{year}{1995}) \bibinfo{pages}{445--451}. \bibinfo{note}{Earth Observation}.
%Type = Article
\bibitem[{Fischer et~al.(2008)Fischer, Birk, Blom, Carli, Carlotti, von Clarmann, Delbouille, Dudhia, Ehhalt, Endemann, Flaud, Gessner, Kleinert, Koopman, Langen, L\'opez-Puertas, Mosner, Nett, Oelhaf, Perron, Remedios, Ridolfi, Stiller, and Zander}]{fischer2008}
\bibinfo{author}{H.~Fischer}, \bibinfo{author}{M.~Birk}, \bibinfo{author}{C.~Blom}, \bibinfo{author}{B.~Carli}, \bibinfo{author}{M.~Carlotti}, \bibinfo{author}{T.~von Clarmann}, \bibinfo{author}{L.~Delbouille}, \bibinfo{author}{A.~Dudhia}, \bibinfo{author}{D.~Ehhalt}, \bibinfo{author}{M.~Endemann}, \bibinfo{author}{J.~M. Flaud}, \bibinfo{author}{R.~Gessner}, \bibinfo{author}{A.~Kleinert}, \bibinfo{author}{R.~Koopman}, \bibinfo{author}{J.~Langen}, \bibinfo{author}{M.~L\'opez-Puertas}, \bibinfo{author}{P.~Mosner}, \bibinfo{author}{H.~Nett}, \bibinfo{author}{H.~Oelhaf}, \bibinfo{author}{G.~Perron}, \bibinfo{author}{J.~Remedios}, \bibinfo{author}{M.~Ridolfi}, \bibinfo{author}{G.~Stiller}, \bibinfo{author}{R.~Zander},
\newblock \bibinfo{title}{Mipas: an instrument for atmospheric and climate research},
\newblock \bibinfo{journal}{Atmospheric Chemistry and Physics} \bibinfo{volume}{8} (\bibinfo{year}{2008}) \bibinfo{pages}{2151--2188}.
%Type = Article
\bibitem[{Dinelli et~al.(2021)Dinelli, Raspollini, Gai, Sgheri, Ridolfi, Ceccherini, Barbara, Zoppetti, Castelli, Papandrea, Pettinari, Dehn, Dudhia, Kiefer, Piro, Flaud, L\'opez-Puertas, Moore, Remedios, and Bianchini}]{dinelli2021}
\bibinfo{author}{B.~M. Dinelli}, \bibinfo{author}{P.~Raspollini}, \bibinfo{author}{M.~Gai}, \bibinfo{author}{L.~Sgheri}, \bibinfo{author}{M.~Ridolfi}, \bibinfo{author}{S.~Ceccherini}, \bibinfo{author}{F.~Barbara}, \bibinfo{author}{N.~Zoppetti}, \bibinfo{author}{E.~Castelli}, \bibinfo{author}{E.~Papandrea}, \bibinfo{author}{P.~Pettinari}, \bibinfo{author}{A.~Dehn}, \bibinfo{author}{A.~Dudhia}, \bibinfo{author}{M.~Kiefer}, \bibinfo{author}{A.~Piro}, \bibinfo{author}{J.-M. Flaud}, \bibinfo{author}{M.~L\'opez-Puertas}, \bibinfo{author}{D.~Moore}, \bibinfo{author}{J.~Remedios}, \bibinfo{author}{M.~Bianchini},
\newblock \bibinfo{title}{The esa mipas/envisat level2-v8 dataset: 10 years of measurements retrieved with orm v8.22},
\newblock \bibinfo{journal}{Atmospheric Measurement Techniques} \bibinfo{volume}{14} (\bibinfo{year}{2021}) \bibinfo{pages}{7975--7998}.
%Type = Article
\bibitem[{Raspollini et~al.(2022)Raspollini, Arnone, Barbara, Bianchini, Carli, Ceccherini, Chipperfield, Dehn, Della~Fera, Dinelli, Dudhia, Flaud, Gai, Kiefer, L\'opez-Puertas, Moore, Piro, Remedios, Ridolfi, Sembhi, Sgheri, and Zoppetti}]{raspollini2022}
\bibinfo{author}{P.~Raspollini}, \bibinfo{author}{E.~Arnone}, \bibinfo{author}{F.~Barbara}, \bibinfo{author}{M.~Bianchini}, \bibinfo{author}{B.~Carli}, \bibinfo{author}{S.~Ceccherini}, \bibinfo{author}{M.~P. Chipperfield}, \bibinfo{author}{A.~Dehn}, \bibinfo{author}{S.~Della~Fera}, \bibinfo{author}{B.~M. Dinelli}, \bibinfo{author}{A.~Dudhia}, \bibinfo{author}{J.-M. Flaud}, \bibinfo{author}{M.~Gai}, \bibinfo{author}{M.~Kiefer}, \bibinfo{author}{M.~L\'opez-Puertas}, \bibinfo{author}{D.~P. Moore}, \bibinfo{author}{A.~Piro}, \bibinfo{author}{J.~J. Remedios}, \bibinfo{author}{M.~Ridolfi}, \bibinfo{author}{H.~Sembhi}, \bibinfo{author}{L.~Sgheri}, \bibinfo{author}{N.~Zoppetti},
\newblock \bibinfo{title}{Level 2 processor and auxiliary data for esa version 8 final full mission analysis of mipas measurements on envisat},
\newblock \bibinfo{journal}{Atmospheric Measurement Techniques} \bibinfo{volume}{15} (\bibinfo{year}{2022}) \bibinfo{pages}{1871--1901}.
%Type = Article
\bibitem[{Boone et~al.(2020)Boone, Bernath, Walker, and McLeod}]{boone2020}
\bibinfo{author}{C.~D. Boone}, \bibinfo{author}{P.~F. Bernath}, \bibinfo{author}{K.~A. Walker}, \bibinfo{author}{S.~D. McLeod},
\newblock \bibinfo{title}{Ace-fts version 3.6 retrieval updates},
\newblock \bibinfo{journal}{Journal of Quantitative Spectroscopy and Radiative Transfer} \bibinfo{volume}{247} (\bibinfo{year}{2020}) \bibinfo{pages}{106939}.
%Type = Article
\bibitem[{Waters et~al.(2022)Waters, Lambert, Livesey, Read, Santee, Froidevaux et~al.}]{waters2022}
\bibinfo{author}{J.~W. Waters}, \bibinfo{author}{A.~Lambert}, \bibinfo{author}{N.~J. Livesey}, \bibinfo{author}{W.~G. Read}, \bibinfo{author}{M.~L. Santee}, \bibinfo{author}{L.~Froidevaux}, et~al.,
\newblock \bibinfo{title}{The earth observing system microwave limb sounder (eos mls) on the aura satellite: A 17-year record of temperature, ozone, and other atmospheric constituents},
\newblock \bibinfo{journal}{Atmospheric Measurement Techniques} \bibinfo{volume}{15} (\bibinfo{year}{2022}) \bibinfo{pages}{1583--1610}.
%Type = Inproceedings
\bibitem[{Riese et~al.(2023)Riese, Preusse, Ungermann, Rhode, Ern, Nogai, Sinnhuber, and Fischer}]{riese2023}
\bibinfo{author}{M.~Riese}, \bibinfo{author}{P.~Preusse}, \bibinfo{author}{J.~Ungermann}, \bibinfo{author}{S.~Rhode}, \bibinfo{author}{M.~Ern}, \bibinfo{author}{K.~Nogai}, \bibinfo{author}{B.-M. Sinnhuber}, \bibinfo{author}{H.~Fischer},
\newblock \bibinfo{title}{Earth explorer 11 mission candidate cairt: Phase 0 system study overview},
\newblock in: \bibinfo{booktitle}{2023 IEEE International Geoscience and Remote Sensing Symposium (IGARSS)}, p. \bibinfo{pages}{10641615}.
%Type = Inproceedings
\bibitem[{Fischer et~al.(2024)Fischer, Riese, Preusse, Ungermann, Rhode, Ern, Nogai, and Sinnhuber}]{fischer2024}
\bibinfo{author}{H.~Fischer}, \bibinfo{author}{M.~Riese}, \bibinfo{author}{P.~Preusse}, \bibinfo{author}{J.~Ungermann}, \bibinfo{author}{S.~Rhode}, \bibinfo{author}{M.~Ern}, \bibinfo{author}{K.~Nogai}, \bibinfo{author}{B.-M. Sinnhuber},
\newblock \bibinfo{title}{The changing-atmosphere infra-red tomography explorer (cairt): A candidate for esa's earth explorer 11},
\newblock in: \bibinfo{booktitle}{EGU General Assembly 2024}, p. \bibinfo{pages}{20214}.
%Type = Article
\bibitem[{Rhode et~al.(2024)Rhode, Preusse, Ungermann, Polichtchouk, Sato, Watanabe, Ern, Nogai, Sinnhuber, and Riese}]{rhode2024}
\bibinfo{author}{S.~Rhode}, \bibinfo{author}{P.~Preusse}, \bibinfo{author}{J.~Ungermann}, \bibinfo{author}{I.~Polichtchouk}, \bibinfo{author}{K.~Sato}, \bibinfo{author}{S.~Watanabe}, \bibinfo{author}{M.~Ern}, \bibinfo{author}{K.~Nogai}, \bibinfo{author}{B.-M. Sinnhuber}, \bibinfo{author}{M.~Riese},
\newblock \bibinfo{title}{Global-scale gravity wave analysis methodology for the esa earth explorer 11 candidate cairt},
\newblock \bibinfo{journal}{Atmospheric Measurement Techniques} \bibinfo{volume}{17} (\bibinfo{year}{2024}) \bibinfo{pages}{5785--5819}.
%Type = Article
\bibitem[{Ridolfi and Sgheri(2014)}]{ridolfi2014}
\bibinfo{author}{M.~Ridolfi}, \bibinfo{author}{L.~Sgheri},
\newblock \bibinfo{title}{Characterization of model errors in the calculation of tangent heights for atmospheric infrared limb measurements},
\newblock \bibinfo{journal}{Atmospheric Measurement Techniques} \bibinfo{volume}{7} (\bibinfo{year}{2014}) \bibinfo{pages}{4117--4122}.
%Type = Book
\bibitem[{Campbell and Wynne(2011)}]{campbell2011}
\bibinfo{author}{J.~B. Campbell}, \bibinfo{author}{R.~H. Wynne}, \bibinfo{title}{Introduction to Remote Sensing}, \bibinfo{publisher}{Guilford Press}, \bibinfo{edition}{5th} edition, \bibinfo{year}{2011}.
%Type = Book
\bibitem[{Vallado(2013)}]{vallado2013}
\bibinfo{author}{D.~A. Vallado}, \bibinfo{title}{Fundamentals of Astrodynamics and Applications}, \bibinfo{publisher}{Microcosm Press and Springer}, \bibinfo{edition}{4th} edition, \bibinfo{year}{2013}.
%Type = Misc
\bibitem[{{European Space Agency}(2020)}]{esa_polar_sun_sync}
\bibinfo{author}{{European Space Agency}}, \bibinfo{title}{Polar and sun-synchronous orbit}, \bibinfo{howpublished}{\url{https://www.esa.int/ESA_Multimedia/Images/2020/03/Polar_and_Sun-synchronous_orbit}}, \bibinfo{year}{2020}. \bibinfo{note}{Accessed: 2025-01-21}.
%Type = Misc
\bibitem[{{National Imagery and Mapping Agency}(2000)}]{wgs84}
\bibinfo{author}{{National Imagery and Mapping Agency}}, \bibinfo{title}{Wgs 84: The world geodetic system 1984}, \bibinfo{howpublished}{\url{https://www.ngs.noaa.gov/WGS84/}}, \bibinfo{year}{2000}. \bibinfo{note}{Accessed: 2025-01-21}.
%Type = Misc
\bibitem[{Pewsey(2021)}]{pewsey_minimum_distance_ellipse}
\bibinfo{author}{M.~Pewsey}, \bibinfo{title}{Minimum distance between ellipse and point}, \bibinfo{howpublished}{\url{https://mpewsey.github.io/2021/11/07/minimum-distance-between-ellipse-and-point.html}}, \bibinfo{year}{2021}. \bibinfo{note}{Accessed: 2025-01-21}.
%Type = Misc
\bibitem[{Nürnberg(2020)}]{nurnberg_distance_to_ellipse}
\bibinfo{author}{R.~Nürnberg}, \bibinfo{title}{Distance from a point to an ellipse}, \bibinfo{howpublished}{\url{https://nurnberg.maths.unitn.it/distance2ellipse.pdf}}, \bibinfo{year}{2020}. \bibinfo{note}{Accessed: 2025-01-21}.
%Type = Misc
\bibitem[{Chatfield(2020)}]{chatfield_distance_to_ellipse}
\bibinfo{author}{C.~Chatfield}, \bibinfo{title}{Simple method for distance to ellipse}, \bibinfo{howpublished}{\url{https://blog.chatfield.io/simple-method-for-distance-to-ellipse}}, \bibinfo{year}{2020}. \bibinfo{note}{Accessed: 2025-01-21}.
%Type = Book
\bibitem[{Born et~al.(1999)Born, Wolf, Bhatia, Clemmow, Gabor, Stokes, Taylor, Wayman, and Wilcock}]{born1999}
\bibinfo{author}{M.~Born}, \bibinfo{author}{E.~Wolf}, \bibinfo{author}{A.~B. Bhatia}, \bibinfo{author}{P.~C. Clemmow}, \bibinfo{author}{D.~Gabor}, \bibinfo{author}{A.~R. Stokes}, \bibinfo{author}{A.~M. Taylor}, \bibinfo{author}{P.~A. Wayman}, \bibinfo{author}{W.~L. Wilcock}, \bibinfo{title}{Principles of Optics: Electromagnetic Theory of Propagation, Interference and Diffraction of Light}, \bibinfo{publisher}{Cambridge University Press}, \bibinfo{edition}{7} edition, \bibinfo{year}{1999}.
%Type = Article
\bibitem[{Edlén(1966)}]{edlen1966}
\bibinfo{author}{B.~Edlén},
\newblock \bibinfo{title}{The refractive index of air},
\newblock \bibinfo{journal}{Metrologia} \bibinfo{volume}{2} (\bibinfo{year}{1966}) \bibinfo{pages}{71--80}.
%Type = Article
\bibitem[{Ciddor(1996)}]{ciddor1996}
\bibinfo{author}{P.~E. Ciddor},
\newblock \bibinfo{title}{Refractive index of air: new equations for the visible and near infrared},
\newblock \bibinfo{journal}{Applied Optics} \bibinfo{volume}{35} (\bibinfo{year}{1996}) \bibinfo{pages}{1566--1573}.
%Type = Article
\bibitem[{Mathar(2004)}]{mathar2004}
\bibinfo{author}{R.~J. Mathar},
\newblock \bibinfo{title}{Calculated refractivity of water vapor and moist air in the atmospheric window at 10 µm},
\newblock \bibinfo{journal}{Applied Optics} \bibinfo{volume}{43} (\bibinfo{year}{2004}) \bibinfo{pages}{928--932}.
%Type = Article
\bibitem[{Mathar(2007)}]{mathar2007}
\bibinfo{author}{R.~J. Mathar},
\newblock \bibinfo{title}{Refractive index of humid air in the infrared: model fits},
\newblock \bibinfo{journal}{Journal of Optics A: Pure and Applied Optics} \bibinfo{volume}{9} (\bibinfo{year}{2007}) \bibinfo{pages}{470}.
%Type = Misc
\bibitem[{Young(2024)}]{young_air_refractivity}
\bibinfo{author}{A.~T. Young}, \bibinfo{title}{Refractivity of air}, \bibinfo{howpublished}{\url{https://aty.sdsu.edu/explain/atmos_refr/air_refr.html}}, \bibinfo{year}{2024}. \bibinfo{note}{Accessed: 2025-01-22}.
%Type = Book
\bibitem[{{National Oceanic and Atmospheric Administration} et~al.(1976){National Oceanic and Atmospheric Administration}, {National Aeronautics and Space Administration}, and {United States Air Force}}]{US1976}
\bibinfo{author}{{National Oceanic and Atmospheric Administration}}, \bibinfo{author}{{National Aeronautics and Space Administration}}, \bibinfo{author}{{United States Air Force}}, \bibinfo{title}{{U.S. Standard Atmosphere, 1976}}, \bibinfo{publisher}{U.S. Government Printing Office}, \bibinfo{address}{Washington, D.C.}, \bibinfo{year}{1976}.
%Type = Article
\bibitem[{Hersbach et~al.(2020)Hersbach, Bell, Berrisford, Hirahara, Horányi, Muñoz-Sabater, Nicolas, Peubey, Radu, Schepers, Simmons, Soci, Abdalla, Abellan, Balsamo, Bechtold, Biavati, Bidlot, Bonavita, De~Chiara, Dahlgren, Dee, Diamantakis, Dragani, Flemming, Forbes, Fuentes, Geer, Haimberger, Healy, Hogan, Hólm, Janisková, Keeley, Laloyaux, Lopez, Lupu, Radnoti, de~Rosnay, Rozum, Vamborg, Villaume, and Thépaut}]{hersbach20}
\bibinfo{author}{H.~Hersbach}, \bibinfo{author}{B.~Bell}, \bibinfo{author}{P.~Berrisford}, \bibinfo{author}{S.~Hirahara}, \bibinfo{author}{A.~Horányi}, \bibinfo{author}{J.~Muñoz-Sabater}, \bibinfo{author}{J.~Nicolas}, \bibinfo{author}{C.~Peubey}, \bibinfo{author}{R.~Radu}, \bibinfo{author}{D.~Schepers}, \bibinfo{author}{A.~Simmons}, \bibinfo{author}{C.~Soci}, \bibinfo{author}{S.~Abdalla}, \bibinfo{author}{X.~Abellan}, \bibinfo{author}{G.~Balsamo}, \bibinfo{author}{P.~Bechtold}, \bibinfo{author}{G.~Biavati}, \bibinfo{author}{J.~Bidlot}, \bibinfo{author}{M.~Bonavita}, \bibinfo{author}{G.~De~Chiara}, \bibinfo{author}{P.~Dahlgren}, \bibinfo{author}{D.~Dee}, \bibinfo{author}{M.~Diamantakis}, \bibinfo{author}{R.~Dragani}, \bibinfo{author}{J.~Flemming}, \bibinfo{author}{R.~Forbes}, \bibinfo{author}{M.~Fuentes}, \bibinfo{author}{A.~Geer}, \bibinfo{author}{L.~Haimberger}, \bibinfo{author}{S.~Healy}, \bibinfo{author}{R.~J. Hogan}, \bibinfo{author}{E.~Hólm}, \bibinfo{author}{M.~Janisková},
  \bibinfo{author}{S.~Keeley}, \bibinfo{author}{P.~Laloyaux}, \bibinfo{author}{P.~Lopez}, \bibinfo{author}{C.~Lupu}, \bibinfo{author}{G.~Radnoti}, \bibinfo{author}{P.~de~Rosnay}, \bibinfo{author}{I.~Rozum}, \bibinfo{author}{F.~Vamborg}, \bibinfo{author}{S.~Villaume}, \bibinfo{author}{J.-N. Thépaut},
\newblock \bibinfo{title}{The era5 global reanalysis},
\newblock \bibinfo{journal}{Quarterly Journal of the Royal Meteorological Society} \bibinfo{volume}{146} (\bibinfo{year}{2020}) \bibinfo{pages}{1999--2049}.
%Type = Misc
\bibitem[{{Copernicus Atmosphere Monitoring Service (CAMS)}(2020)}]{egg2020}
\bibinfo{author}{{Copernicus Atmosphere Monitoring Service (CAMS)}}, \bibinfo{title}{Cams global greenhouse gas reanalysis (egg4)}, \bibinfo{howpublished}{\url{https://atmosphere.copernicus.eu/cams-global-greenhouse-gas-reanalysis}}, \bibinfo{year}{2020}. \bibinfo{note}{Accessed: 2024-09-23}.
%Type = Article
\bibitem[{Remedios et~al.(2007)Remedios, Leigh, Waterfall, Moore, Sembhi, Parkes, Greenhough, Chipperfield, and Hauglustaine}]{remedios07}
\bibinfo{author}{J.~J. Remedios}, \bibinfo{author}{R.~J. Leigh}, \bibinfo{author}{A.~M. Waterfall}, \bibinfo{author}{D.~P. Moore}, \bibinfo{author}{H.~Sembhi}, \bibinfo{author}{I.~Parkes}, \bibinfo{author}{J.~Greenhough}, \bibinfo{author}{M.~P. Chipperfield}, \bibinfo{author}{D.~Hauglustaine},
\newblock \bibinfo{title}{Mipas reference atmospheres and comparisons to v4.61/v4.62 mipas level 2 geophysical data sets},
\newblock \bibinfo{journal}{Atmospheric Chemistry and Physics Discussions} \bibinfo{volume}{7} (\bibinfo{year}{2007}) \bibinfo{pages}{9973--10017}.
%Type = Article
\bibitem[{Raspollini et~al.(2013)Raspollini, Carli, Carlotti, Ceccherini, Dehn, Dinelli, Dudhia, Flaud, L\'opez-Puertas, Niro, Remedios, Ridolfi, Sembhi, Sgheri, and von Clarmann}]{raspollini2013}
\bibinfo{author}{P.~Raspollini}, \bibinfo{author}{B.~Carli}, \bibinfo{author}{M.~Carlotti}, \bibinfo{author}{S.~Ceccherini}, \bibinfo{author}{A.~Dehn}, \bibinfo{author}{B.~M. Dinelli}, \bibinfo{author}{A.~Dudhia}, \bibinfo{author}{J.-M. Flaud}, \bibinfo{author}{M.~L\'opez-Puertas}, \bibinfo{author}{F.~Niro}, \bibinfo{author}{J.~J. Remedios}, \bibinfo{author}{M.~Ridolfi}, \bibinfo{author}{H.~Sembhi}, \bibinfo{author}{L.~Sgheri}, \bibinfo{author}{T.~von Clarmann},
\newblock \bibinfo{title}{Ten years of mipas measurements with esa level 2 processor v6; part 1: Retrieval algorithm and diagnostics of the products},
\newblock \bibinfo{journal}{Atmospheric Measurement Techniques} \bibinfo{volume}{6} (\bibinfo{year}{2013}) \bibinfo{pages}{2419--2439}.
%Type = Article
\bibitem[{Kiefer et~al.(2010)Kiefer, von Clarmann, Grabowski, Stiller, Glatthor, Höpfner, Kellmann, Linden, and Mengistu~Tsidu}]{kiefer2010}
\bibinfo{author}{M.~Kiefer}, \bibinfo{author}{T.~von Clarmann}, \bibinfo{author}{U.~Grabowski}, \bibinfo{author}{G.~P. Stiller}, \bibinfo{author}{N.~Glatthor}, \bibinfo{author}{M.~Höpfner}, \bibinfo{author}{S.~Kellmann}, \bibinfo{author}{A.~Linden}, \bibinfo{author}{G.~Mengistu~Tsidu},
\newblock \bibinfo{title}{Impact of temperature field inhomogeneities on the retrieval of atmospheric species from mipas ir limb emission spectra},
\newblock \bibinfo{journal}{Atmospheric Measurement Techniques} \bibinfo{volume}{3} (\bibinfo{year}{2010}) \bibinfo{pages}{1487--1507}.
%Type = Misc
\bibitem[{Schneider et~al.(2005)Schneider, Bechtold, Martin et~al.}]{kopra}
\bibinfo{author}{M.~Schneider}, \bibinfo{author}{P.~Bechtold}, \bibinfo{author}{J.~Martin}, et~al., \bibinfo{title}{Kopra - a radiative transfer model for atmospheric applications}, \bibinfo{year}{2005}. \bibinfo{note}{Version 5.1}.
%Type = Article
\bibitem[{Godson(1953)}]{godson1953}
\bibinfo{author}{W.~L. Godson},
\newblock \bibinfo{title}{The evaluation of infra‐red radiative fluxes due to atmospheric water vapour},
\newblock \bibinfo{journal}{Quarterly Journal of the Royal Meteorological Society} \bibinfo{volume}{79} (\bibinfo{year}{1953}) \bibinfo{pages}{367--379}.
%Type = Article
\bibitem[{Savitzky and Golay(1964)}]{savitzky1964}
\bibinfo{author}{A.~Savitzky}, \bibinfo{author}{M.~J.~E. Golay},
\newblock \bibinfo{title}{Smoothing and differentiation of data by simplified least squares procedures},
\newblock \bibinfo{journal}{Analytical Chemistry} \bibinfo{volume}{36} (\bibinfo{year}{1964}) \bibinfo{pages}{1627--1639}.

\end{thebibliography}

\end{document}